%% file: lux.tex
\documentstyle[aps,epsf,amssymb]{revtex}
\tightenlines
\begin{document}

\title
 {The RANLUX generator: resonances in a random walk test.}

\author{Lev N. Shchur\cite{lns}  and Paolo Butera\cite{pb} }
\address
{Landau Institute for Theoretical Physics, 142432 Chernogolovka, Russia
\\and\\
Istituto Nazionale di Fisica Nucleare, Dipartimento di Fisica,
Universit\`a di Milano, \\
Via Celoria 16, 20133 Milano, Italy}

\maketitle

\begin{abstract}
Using a recently proposed directed random walk test, we systematically
investigate the popular random number generator RANLUX developed by
L\"uscher and implemented by James. 
We confirm the good quality of this generator with the recommended
luxury level. At a smaller luxury level  (for instance equal to 1)
resonances are observed in the random walk test.
We also find that the lagged Fibonacci and
Subtract-with-Carry recipes exhibit similar failures
in the random walk test. A revised analysis of the corresponding
dynamical systems leads to the observation of resonances in
the eigenvalues of Jacobi matrix.
\end{abstract}


\widetext

\section{Introduction}

 Large-scale Monte Carlo simulations need good quality random numbers
\cite{BiHe}. The half-century long history of computer simulations shows
successes and failures of many algorithms for pseudo-random number (PRN)
generation
\cite{knuth}.

 Usually, whenever a new test of randomness is proposed some
algorithm for generating PRN's becomes obsolete.
A somewhat different point of view is advocated here: using a
recently proposed test we argue that all algorithms  using feedbacks
(or lags) have more or less the same defects and belong to the same
"universality class of badness". This class of generators, however,
should not be rejected and rather improved versions of them can still be
considered reasonably safe (see the final section).

 One of the most widely used PRN generators
of this class is the shift register (SR) generator \cite{golomb}.
 Correlations in SR generators were pointed out by Compagner
\cite{At} but the warning was ignored by computational physicists who
usually stick to very practical recipes\cite{flue}.

 It has been observed later by Ferrenberg, Landau and Wong \cite{flw} that
the statistical simulations of 2D Ising model by the Wolff
single cluster algorithm are very sensitive to the defects of the
Kirkpatrick-Stoll generator \cite{KS} (the SR generator with
$(p,q)=(250,103)$).

 Motivated by this difficulty, L\"uscher\cite{lush} and James\cite{james}
have developed a random number generator called RANLUX
which is based on the subtract-with-carry (SWC) recipe advocated
by Marsaglia and Zaman\cite{mz}.

 We review here some properties of the generator RANLUX using the random
walk test recently developed in \cite{SHB} by Bl\"ote, Heringa and
one of the authors.

 First, we find analytically that the deviations computed using the
random walk test are the same for SWC and the lagged
Fibonacci recipes. This finding is in contrast to the common
belief  that the SWC generator gives better quality random numbers than
recipes which use the exclusive OR operation (like the SR generators)
or addition (like the lagged Fibonacci generator). 
Note, that the RANLUX generator with
parameter of luxury equal to $0$ is equivalent to the SWC generator.

 Second, we have found numerically, that correlations among random numbers
on distances associated with the lag values are still visible for luxury
level parameter equal to $1$. However at luxury level equal to $2$ the
correlations lie on the boundary of observed deviations (and this is in
agreement with the L\"uscher and James results). 

 Therefore, our third result, supporting the author's expectations,
is that the RANLUX generator with luxury levels higher than $2$ is still
safe in runs using less than $10^{15}$ random numbers.

 Next, we also revise the dynamical system description given by L\"uscher  in his
original paper. Namely, we argue that the correlations responsible for the
deviations found in the simulation of the two-dimensional Ising model 
and of the random
walks are not due to the correlation of trajectories in the corresponding
dynamical systems. These deviations are rather connected with the
correlations between random numbers, on distances equal to
the lag values, as we show using the random walk test. 

 We compute the full spectrum of eigenvalues for the RANLUX generator and
for the lagged Fibonacci generator with different lags. Our fifth result
is the observation of resonances in the eigenvalues, reflecting the
complexity of the
phase space in the area preserving (Hamiltonian) dynamical systems.

 The paper is organized is follows. In section~\ref{sec-DN}
we recall the definitions
of random number generators used here - Subtract-with-Carry (SWC), lagged
Fibonacci (LF) and Shift register (SR), and review the random walk test.
In section~\ref{sec-RW}
we apply the random walk test to the RANLUX generator and surprisingly
find that the SWC generator obeys the same analytical solutions
as the LF generator.
Then we analyse how the correlations in random walks depend on the luxury
levels of the RANLUX generator. In section~\ref{sec-DS}
we reanalyse the dynamical system
approach to  RANLUX generator and find resonances in the eigenvalues of
the Jacobi matrix associated with the complex structure of the manifolds.
Finally, in section~\ref{sec-Con} we discuss the attempts to improve the
lagged random number generators.

\section{ Definitions and notations}
\label{sec-DN}

 We recall here the definitions of several random number generators
which have the common feature of using lag algorithms.
We also recall the definition of the random walk test.

\subsection{Marsaglia-Zaman recipe and RANLUX generator}

The  PRN generation algorithm of Marsaglia-Zaman\cite{mz} is defined
by a recursion relation involving three fixed positive integers
$b,r,s$ where $b$ is called the base and $r > s$ are called the lags.
This algorithm is usually known as SWC generator.
 Given the first $r$ PRN's $x_0, x_1, ...x_{r-1}$ and the "carry bit"
$c_{r-1}$, the n-th PRN $(n \ge r)$ is given by

\begin{equation}
\begin{array}{l}
 x_n = (x_{n-s} - x_{n-r} - c_{n-1}) \;  {\rm mod} \; b \\
 c_n = 0 \;\;\;\;  if \;\; x_{n-s} - x_{n-r} \ge 0   \\
{\rm and}    \\
c_n = 1 \;\;\;\; {\rm otherwise}
\label{SWC}
\end{array}
\end{equation}

The maximum possible period of this generator is $M= b^r-b^s+1$ and it is
attained when $M$ is prime and $b$ is a primitive root modulo $M$.

In order to improve the properties of this algorithm which fails to
pass some
correlation tests,  L\"uscher proposed \cite{lush} to discard some of the
PRN's produced by the MZ recursion and to use only the remaining ones.
 James has then defined \cite{james} four levels of rejection
(called  "luxury levels"),
characterized by an integer $p\ge 24$
in which the generator produces 24 PRN's, then discards the successive $p-24$
and so on.
Clearly the value $p=24$ reproduces the original MZ recipe where all PRN
are kept  and it is called luxury level zero. The values
$p=48,97,223,389$ define the luxury  levels $1,2,3,4$ respectively.
It has been suggested \cite{lush} that level 3 has a good chance of
being optimal.

\subsection{Lagged Fibonacci generator}

The Fibonacci lagged generator (LF) is defined by the recursion relation
\begin{equation}
x_n = (x_{n-s} + x_{n-r}) \; {\rm mod} \;  2^w
\label{LF}
\end{equation}

where again $s$ and $r>s$ are the lags and $2^w$ is the base.
Brent gave conditions \cite{brent}, under which the above relation
generates a sequence of PRN's having the maximum possible period
$T=2^{w-1}(2^r-1)$. ($T$ is simply the product of the reduced computer
word length and of the number of possible initial seeds.
The latter is the number of possible $r$ bit seeds excluding zero.)

\subsection{Shift Register generator}

The generalized feedback shift register (SR) method is
defined by the recursion rule \cite{golomb}

\begin{equation}
x_n = x_{n-s} \oplus x_{n-r}
\label{SR}
\end{equation}
where $\oplus$ denotes the {\it exclusive-or} operation
(bitwise sum modulo 2), $s$ and $r>s$ are the lags.
This sequence has maximum period $T=2^r-1$ if its
characteristic polynomial is irreducible (for details, see Chapter II of
Golomb book \cite{golomb}).

\subsection{Random Walk Test}

Let us consider the one-dimensional directed random walk model \cite{SHB}:
a walker starts at some site of an one-dimensional lattice and, at
discrete times $i$, either he takes a step in a  fixed  direction with a
probability $\mu_i$ or he stops with a probability $1-\mu_i$.
 In the latter case a new walk begins. The probability of a walk
with length $n$ is then
$$
 P(n) = \left(\prod_{i=1}^{n-1} \mu_i\right) \; (1-\mu_n).
$$

 In the case in which all probabilities are equal $\mu_i=\mu$, the
the probability for a walk of length $n$ is
\begin{equation}
P(n)=\mu^{n-1}\; (1-\mu )
\label{Prob-p}
\end{equation}
and the mean walk length is given by
\begin{equation}
\langle n \rangle =\frac{1}{1-\mu }.
\label{Mean-p}
\end{equation}

Comparing $\mu$ with $tanh(J/k_BT)$ it immediately follows
that the mean walk size diverges as
$\mu\rightarrow 1$ in the same way as the mean cluster size diverges
as the temperature $T\rightarrow 0$ in the 1D Ising model \cite{SHB}.

\section{Random walk in a correlated environment}
\label{sec-RW}

The production rules for the the SR and LF generators  with the pair
$(p,q)$ of lags give rise to correlations on distances $l_c=ip+jq$,
where $i$ and $j$ are integers
\cite{At,SHB}. We can treat these correlations as some kind of
impurities, placed on the bonds of 1D lattice and thus influencing the
walk probabilities $\mu_i$.

In some cases, it is possible to get exact analytical results for the
deviations of walk probabilities  from their values in the pure case.

In addition, correlations in the PRN sequences can be nicely detected by
simulation of this model on the computer
and by comparing the observed  frequency $P^*(n)$ of walks of length $n$
with their expected (pure) probability
$P(n)$. In what follows we will plot the measured deviations
\begin{equation}
 \delta P(n) \equiv \frac{P^*(n)}{P(n)} -1
\label{dP}
\end{equation}
versus $n$, for various PRN generators and briefly comment on why they
occur.

For all PRN generators we will consider here, biases are brought out in the
directed random walk model by the obvious fact that the end of a walk and the beginning of a
new one are strictly correlated: as a result probability deviations will occur
at walk lengths related to the lag values in the production rule.
 To be definite, suppose for example that in our PRN sequence the n-th term
$x_n$ is generated from  $x_{n-r}$ and $x_{n-s}$ with $r>s$
and let us assume for simplicity that the stepping probability in our model has
the value $\mu=1/2$. In other words the ideal walker would
proceed after tossing a good coin, while the real walker will have to resort to
our deterministic PRN generator with lags $r$ and $s$.
Let us now suppose that the real walker stops after $k$
steps, so that $x_k > \mu$.
 Since the production rule  gives $x_k$ in terms of
$x_{k-r}$ and $x_{k-s}$, both of which have to be less than $\mu$,
it follows that  the probability of the k-th step
of the real walker is biased by his previous history.

The first study of this new correlation test in the case of the shift
register and lagged Fibonacci generators has been performed in
\cite{SHB}.

\subsection{Random walk and luxury levels}

 Here we shall test the RANLUX generator with different levels of luxury
in the simulation of the directed random walk model.

 To begin let us choose luxury level $0$ and step probability $\mu=31/32$. In
Fig.~1 we have plotted versus $n$ the deviation $\delta P(n)$
defined by (\ref{dP}) of the frequency of the walks of length $n$ from their
expected probability. We notice strong signals for $n_1=r=24$ and
$n_2=2r=48$. A sizable deviation is also clearly visible for all walk
lengths exceeding the largest lag $r$.

We should stress that our choice of the value of the step probability has
no special meaning and that our results depend smoothly on $\mu$.

\begin{figure}
\epsfxsize=250pt
\epsffile{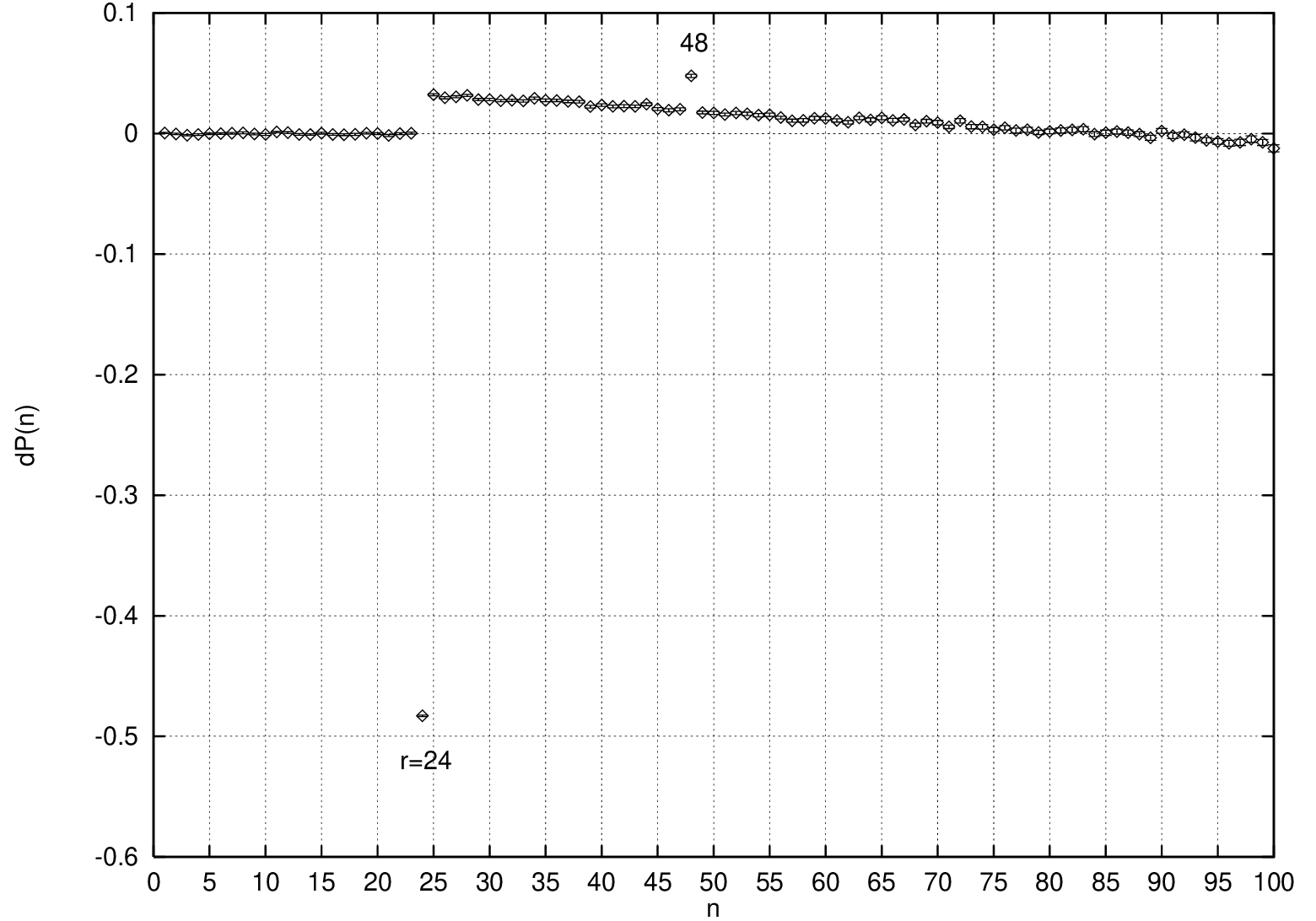}
\vskip 5mm
\caption{Deviation $\delta P$ of the probability of a walk length
$n$ from the value for uncorrelated random numbers versus walk length.
We have used the RANLUX random number generator with lag values $r=24$ 
and $s=10$
and luxury level $0$. The step probability is $\mu=31/32=0.968750$.
The result of averaging over $10^8$ walks is shown.}
\label{pb1}
\end{figure}

 One should notice the qualitative similarity of the deviations in
Fig.~\ref{pb1} and
Fig.~\ref{pb6}, which refers to the case of a lagged Fibonacci generator.
Moreover, it could be shown analytically
that the deviations at $n=r$ and at $n=i$, satisfying
\begin{equation}
r<i<\left\{\begin{array}{lll}
r+s& \mathrm{if}&s<\frac{r}{2}\\
2r-s& \mathrm{if}&s>\frac{r}{2}\\
\end{array}
\right.
\label{ineq}
\end{equation}
coincide for the lagged Fibonacci and the SWC generators. We recall
that the RANLUX generator with luxury level $0$ is equivalent to the  SWC
generator. It also could be shown, that
the mean value of the deviation for the SWC generators at $n=r$ is equal
to that for the lagged Fibonacci generator obtained in \cite{SHB}
\begin{equation}
\delta P(r)=\frac{1-2\mu}{2\mu}
\label{dPr}
\end{equation}
and that the mean value of the deviation at $n=r+1$ is
\begin{equation}
\delta P(r+1)=\frac{(3\mu-1)^2}{4\mu^4}-1.
\label{dPr1}
\end{equation}

 In order to check that point, in Table~\ref{Table1} we have presented
a comparison of the observed values of
the deviations obtained using either the RANLUX generator with luxury
level $0$ or the lagged Fibonacci generator for two values of the step
probability with the expected values given by (\ref{dPr}) and
(\ref{dPr1}). All data agree within the statistical errors.

 In contrast to common belief this result shows that there are no
differences in the accuracy of the data obtained using the LF
and the SWC generators, at least as far as the  cluster
formation processes or the random walks are simulated.

\begin{figure}
\epsfxsize=250pt
\epsffile{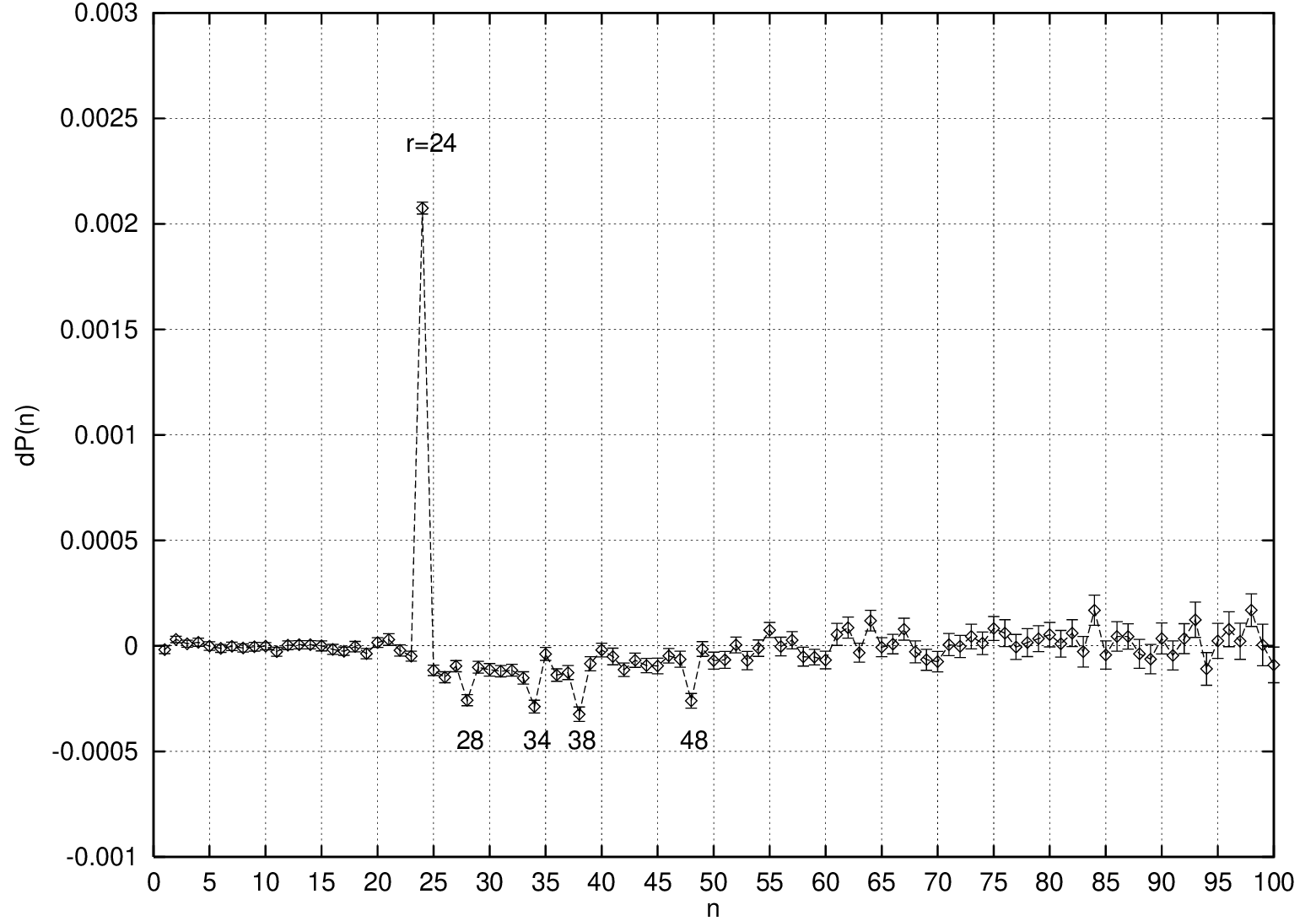}
\vskip 5mm
\caption{Deviation $\delta P$ of the probability of a walk length
$n$ from the value for uncorrelated random numbers versus walk length.
We have used the RANLUX random number generator with lag values $r=24$ and $s=10$
and luxury level $1$. The step probability is $\mu=31/32=0.968750$.
The result of averaging over $10^{11}$ walks is shown.}
\label{pb4}
\end{figure}

 The next luxury level of L\"uscher and James RANLUX is $1$. In this case
only the first $r=24$ random numbers are used out of each generated set of
$p=48$.
The observed deviations (\ref{dP}) are plotted in Fig.~\ref{pb4}.
Clearly, a deviation at $n=r=24$ still remains, but has opposite sign
and a much smaller size in comparison with that observed for the
luxury level $0$.

Also at the values $n=28,34,38,48$, all of which are linear combinations
of the lags with small integer coefficients, smaller deviations remain
clearly visible.

We conclude that, at the luxury levels $0$ and $1$, the deviations have
the same qualitative nature, the only difference being that the absolute
value of the deviations is depressed for the higher levels of luxury.

\begin{figure}
\epsfxsize=250pt
\epsffile{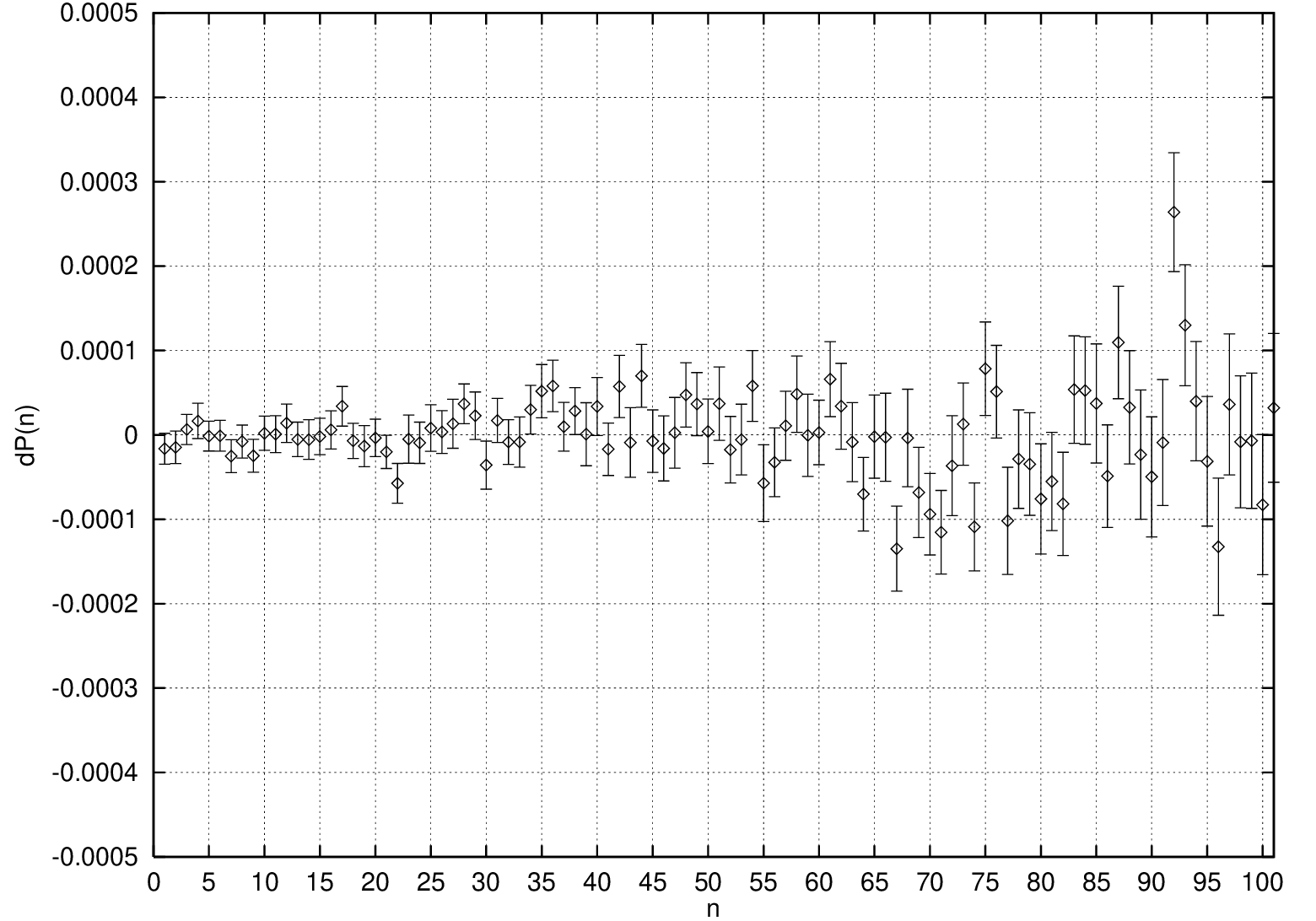}
\vskip 5mm
\caption{Deviation $\delta P$ of the probability of a walk length
$n$ from the value for uncorrelated random numbers versus walk length.
We have used the RANLUX random number generator with lag values $r=24$ and $s=10$
and luxury level $2$. The step probability is $\mu=31/32=0.968750$.
The result of averaging over $10^{11}$ walks is shown.}
\label{pb5}
\end{figure}

\begin{figure}
\epsfxsize=250pt
\epsffile{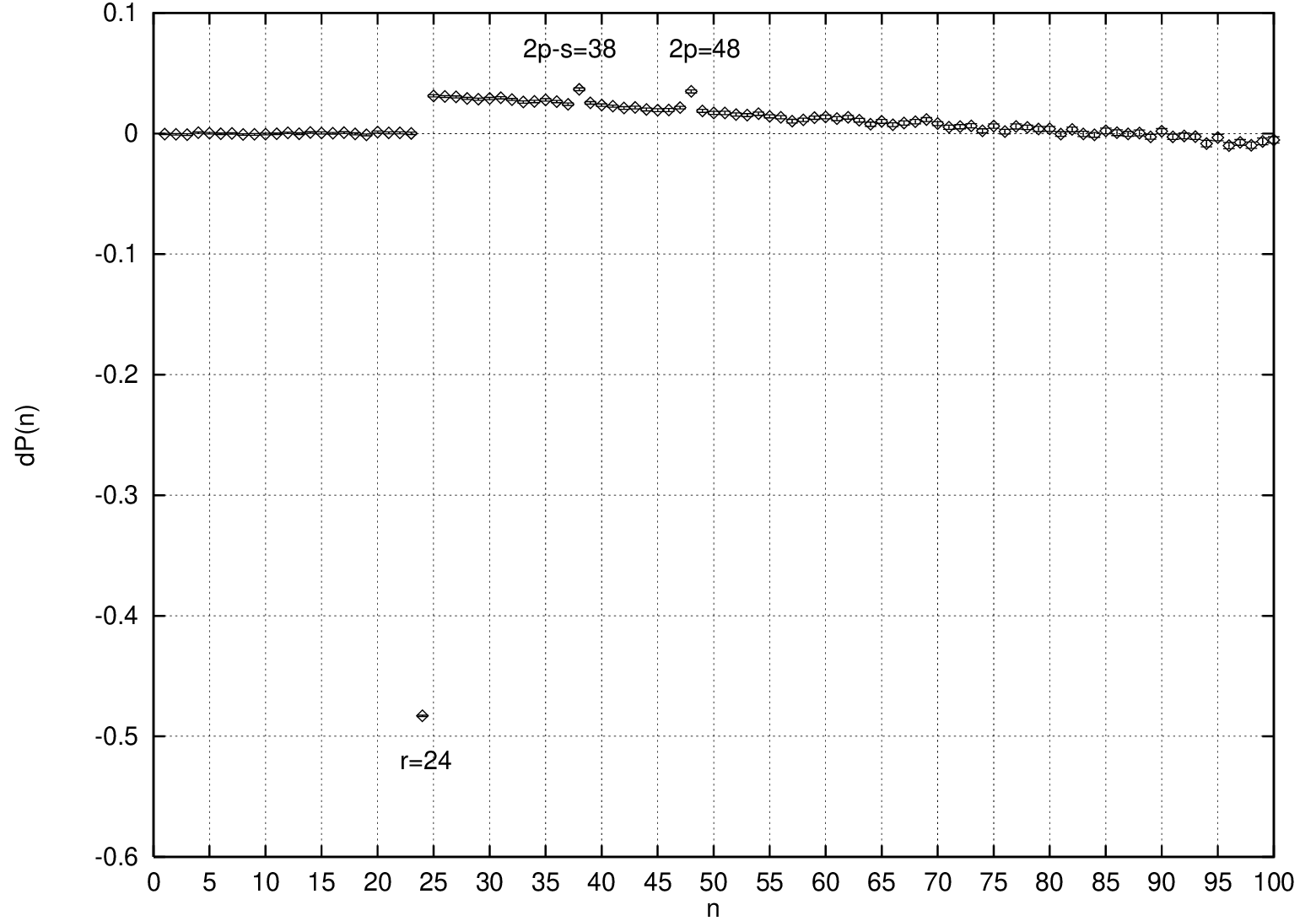}
\vskip 5mm
\caption{Deviation $\delta P$ of the probability of a walk length
$n$ from the value for uncorrelated random numbers versus walk length.
We have used the lagged Fibonacci generator with lag values $r=24$ and $s=10$.
The step probability is $\mu=31/32=0.968750$.
The result of averaging over $10^8$ walks is shown.}
\label{pb6}
\end{figure}

This remark is confirmed by Fig.~\ref{pb5} where we have plotted the
observed deviations for the case of luxury level $2$, and the
visible deviations are not larger than the statistical errors.
Notice that in this case the average length of the walk is 32 and for the
total number of $10^{11}$ walks approximately $10^{13}$ random numbers were
used. We could extrapolate our result in like in Ref.~\cite{lh} and we
expect that deviations will be visible even for the luxury level 2
if $\approx 10^{15}$ random numbers are used. 

\subsection{Lagged Fibonacci with luxuries}

In Fig.~\ref{pb6} we present the results of a simulation using the plain
lagged Fibonacci generator with lags $r=24$ and $s=10$. The plot shows
features completely similar to those of Fig.~\ref{pb1} obtained with
RANLUX at luxury level $0$.

\begin{figure}
\epsfxsize=250pt
\epsffile{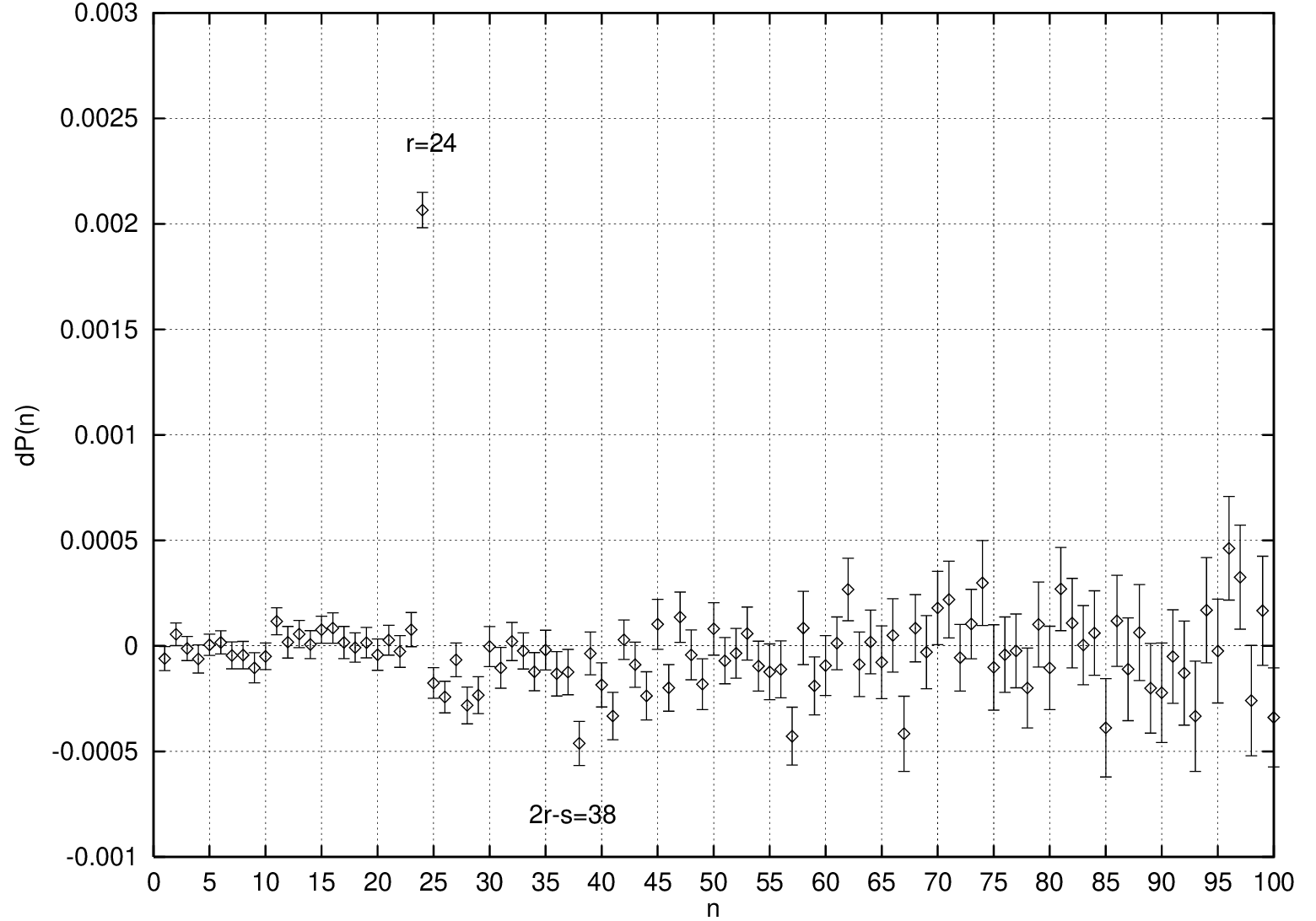}
\vskip 5mm
\caption{Deviation $\delta P$ of the probability of a walk length
$n$ from the value for uncorrelated random numbers versus walk length.
We have used the lagged Fibonacci  generator with lag values $r=24$ and $s=10$ and luxury level $1$.
The step probability is $\mu=31/32=0.968750$.
The result of averaging over $10^{10}$ walks is shown.}
\label{pb7}
\end{figure}

We can now extend to the lagged Fibonacci generator (LF) the procedure of
decorrelation which discards $24$ PRN's out of each generated set of
$48$, and call it following L\"uscher \cite{lush} and James
\cite{james} luxury level $1$. In Fig.~\ref{pb7} we have reported the
results of a simulation using the LF generator with lags $r=24$, $s=10$ at
luxury level $1$. The results again show no qualitative
difference from those reported in Fig.~\ref{pb4} obtained when RANLUX at
the same luxury level was used: a strong deviation is observed when the
walk length equals the value of the largest lag $r$ and smaller but
visible deviations occur for any $n>r$.

\begin{figure}
\epsfxsize=250pt
\epsffile{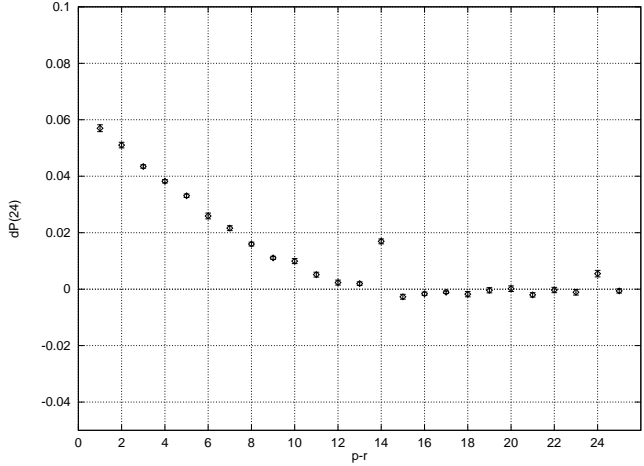}
\vskip 5mm
\caption{Lagged Fibonacci  generator with lags $r=24$ and $s=10$ and $\mu=15/16$.
The figure shows the deviation $\delta P(r)$ as a function of $p-r$ on
the way from luxury level $0$ ($p-r=0$) to luxury level $1$ ($p-r=24$).}
\label{way01}
\end{figure}

In order to understand how the decorrelation procedure works we also have
studied intermediate luxury levels obtained by varying $p$ between $24$ and
$48$ in steps of one. In Fig.~\ref{way01} we have plotted
$\delta P(24)$ versus $p-r$. The value of $\delta P(24)$ decreases rapidly
but peaks are still evident at $p-r=10,14,24$ indicating the persistence
of an interplay between $p$ and the lags $r$ and $s$.

\section{Dynamical System}
\label{sec-DS}

L\"uscher justified his proposal of skipping some $p-r$ numbers using the
language of dynamical systems\cite{lush}.

Let us then briefly recall here a few well known results from the
dynamical system theory.

The production rules (\ref{SWC}),(\ref{LF}) and (\ref{SR}) can be
rewritten \cite{lush} in the matrix form

\begin{equation}
{\bf v}_{n+1}={\bf L}\; {\bf v}_{n},
\label{vect}
\end{equation}
where the rule of matrix "multiplication" includes the "$\oplus$"
operation for the shift register  generator, the usual "$+$" operation
for the lagged Fibonacci
and the "$-$" operation for the SWC generators,
and ${\bf v}_n=(x_{n-r+1},x_{n-r+2},...,x_{n-1},x_n)$ is an $r$-component
vector.

The simplest example is the Fibonacci sequence
\begin{equation}
x_n=x_{n-1} + x_{n-2},
\label{sr2}
\end{equation}
which can be written in matrix form
\begin{equation}
\left(
\begin{array}{l}
x_{n} \\ x_{n-1}
\end{array} \right)
=
\left(
\begin{array}{cc}
1 & 1 \\
1 & 0
\end{array} \right)
\left(
\begin{array}{l}
x_{n-1} \\ x_{n-2}
\end{array} \right).
\label{sr2-m}
\end{equation}
This shows that the dynamical system, corresponding to the Fibonacci
random number
generator with lags $r=2$ and $s=1$ is the famous Arnold's "cat map",
acting on the unit torus, obtained by rewriting (\ref{vect}) in the form
\begin{equation}
{\bf v}_{n+2}={\cal M}\; {\bf v}_{n}.
\label{Mon}
\end{equation}

The matrix ${\cal M}={\bf L}^r$ is called the monodromy matrix,
\begin{equation}
{\cal M}=
\left(
\begin{array}{cc}
2 & 1 \\
1 & 1
\end{array} \right).
\label{cat}
\end{equation}

The eigenvectors $(\gamma^{-2},1)$ and $(\gamma^2-1,1)$
of $\cal M$ define respectively the stable and the unstable directions
of the origin, which is an unstable fixed point. The corresponding
eigenvalues are
\begin{equation}
\left(
\begin{array}{l}
\lambda_1 \\
\lambda_2
\end{array} \right)
=
\left(
\begin{array}{l}
\gamma^2 \\
\gamma^{-2}
\end{array} \right)
\approx
\left(
\begin{array}{l}
2.618034... \\
.381966...
\end{array} \right),
\label{cat-eig}
\end{equation}
and $\gamma=(1+\sqrt{5})/2$ is the golden mean.

\begin{figure}
\epsfysize=220pt
\epsfxsize=250pt
\epsffile{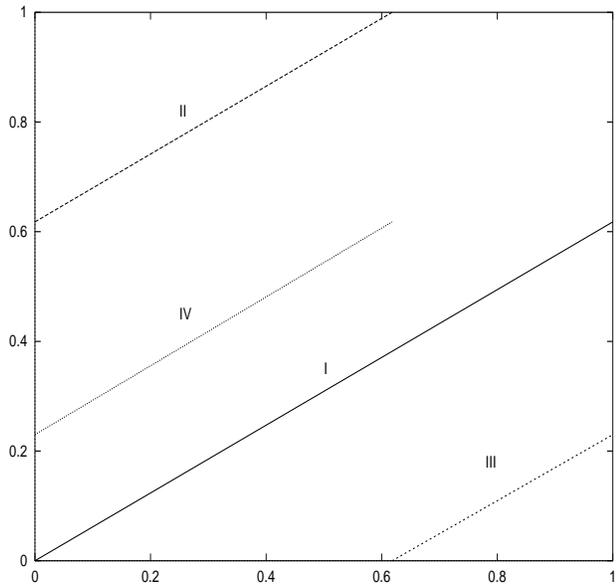}
\vskip 5mm
\caption{Fibonacci sequence on torus: Arnold's cat map.}
\label{cat-t}
\end{figure}

Fig.~\ref{cat-t} demonstrates the idea of hyperbolicity of the "cat map".
The points on the part~$I$ of the unstable manifold are mapped under
a single  application of the transformation (\ref{cat}) onto the four
parts (in order, denoted by I, II, III and IV) of the same unstable
manifold. Clearly, Fig.~\ref{cat-t}
shows a trajectory winding on the torus with the golden angle
$tan\; \phi =(\sqrt{5}-1)/2$.
The distance between neighbor points along the unstable manifold
is enlarged by a factor $\lambda_1$.
In order to imagine the complexity of the trajectories, one needs
to add the set of transversal lines representing the stable manifold.
Along these lines the distances are contracted by a factor $\lambda_2$.
Thus, according to the Poincar\'e theory, there is an infinite set
of periodic points with zero measure in this system  (we address
interested readers to the figures on page 170 of Ref.~\cite{tabor}.)

It is known \cite{arnold} that the "cat map" is an Anosov system,
which implies the following properties of randomness \cite{tabor}:
global instability (the Lyapunov exponent, which
is the logarithm of $\lambda_1$ from Eq.~\ref{cat-eig}, is positive),
positivity of Kolmogorov-Sinai entropy (the trajectories are
locally divergent), mixing (existence of equilibrium state for the
distribution function) and, finally, ergodicity (the equivalence of
averaging over time and over space). Nevertheless, the "cat map" doesn't
have the strongest property of randomness, namely, it is not a Bernoulli
system, in other words (M. Tabor in Ref.~\cite{tabor}, page 174) it is not
a system "whose motion is as random as a fair coin toss".

In fact, the lagged Fibonacci generators (and the Marsaglia-Zaman
SWC generators as well) are an $r$-dimensional generalization of
a "cat map". A regular way to extend randomness properties to higher
dimensions is not known. Nevertheless, in general,  one should not expect
from the generalized system stronger properties of randomness than in the
case of a low dimensional system. In fact, practically it is sufficient to
analyze the eigenvalues of the corresponding monodromy matrix to be sure
that a system is locally hyperbolic in any point of phase space (to have, at
least, the properties of the weakly Anosov systems \cite{MacKay}).

The suggestion of L\"uscher to improve the quality of the random numbers
generated by
rule (\ref{vect}) is based on the observation, that since the
absolute value of the largest eigenvalue of the matrix $L$ in (\ref{vect})
is greater than unity, we can construct a monodromy matrix
${\cal M}={\bf L}^p$ with a relatively large modulus of the
largest eigenvalue, whose logarithm is the Lyapunov exponent of
the corresponding dynamical system on the $r$-dimensional torus.

Indeed, his Fig.~1 from Ref. \cite{lush} shows clearly the process
of decorrelation (loos of memory). If we apply the monodromy
matrix $\cal M$ to the shortest representable vector which has elements of
order of $2^{-m}$, the length of our vector will become of order of unity
after a number of iterations $n$ such that
\begin{equation}
2^{-m}\; \lambda^n \approx \; 1.
\label{approx}
\end{equation}
 Knowing L\"uscher result for $n=16$ and considering the case of $24$-bit
representation for the random numbers ($m=24$), as in his paper,
from (\ref{approx}) we could estimate $\lambda$ as
$\lambda\approx 2^{24/16}\approx 2.828...$ in a good agreement with his
direct calculation $\lambda=2.746...$.

This means, by L\"uscher, that the decorrelation time depends
on the number of bits in the random number representation
\begin{equation}
n\approx m\;\frac{\log 2}{\log \lambda}
\label{contr}
\end{equation}
and leads to the prediction of an infinite decorrelation time in the
limit $m\rightarrow\infty$. We could use the same arguments for
the case of a "cat map" and would arrive to a contradiction with Arnold's
result, that the "cat map" is a C-system with the strong ergodic
properties cited above.

This means, that the values of the Lyapunov exponent are not directly
connected with the process of decorrelation in random number sequences.
We can support our idea considering LCG generator which are known to have
a lattice structure and introducing luxury levels in this case. It is
clear, that the lattice structure will not disappear -
increasing the luxury levels would enlarge eigenvalues, but the
lattice structure would still persist.

 Really, as we have demonstrated using the random walk test,
the situation is more complicated and we need to consider also
the correlations in the production rules on distances equal to the lags
and to their linear combinations.  This effect is clearly visible in
Fig.~\ref{way01}, where one can see strong resonances at lag values
$r=24$ and $s=10$, whereas the eigenvalues are increasing smoothly
as $\lambda_1^p$ with $p$.

 It should be noted, that the deviations at $\delta P(r)$ do not vanish
smoothly on the way from one luxury level to another, but they
oscillate. We hope, nevertheless, that the amplitude of these
oscillations vanishes.

\subsection{Eigenvalue spectrum}

 Each state of the random number generators of interest is described by
a set of $r$ random numbers $v^0_i, \; (i=1,2,...,r)$. After
$r$ applications of any of the rules (\ref{SWC}),(\ref{LF}) and
(\ref{SR}), we will have completely refreshed the set of $r$ random
number $u^0_i, \; (i=1,2,...,r)$. This defines the mapping of
unit $r$-dimensional cube onto itself\cite{lush}.

We could construct numerically the Jacobi matrix
\cite{arnold,ym2}, as the matrix of the variations of the final set
$u_i=u^0_i+\delta u_i$ due to the variations of the initial set
$v_i=v^0_i+\delta v_i$:

\begin{equation}
J_{ij}=\frac{\delta u_i}{\delta v_j},
\label{Yac}
\end{equation}
which reflects the behavior of neighbor trajectories
for all components of the set $u_i$.

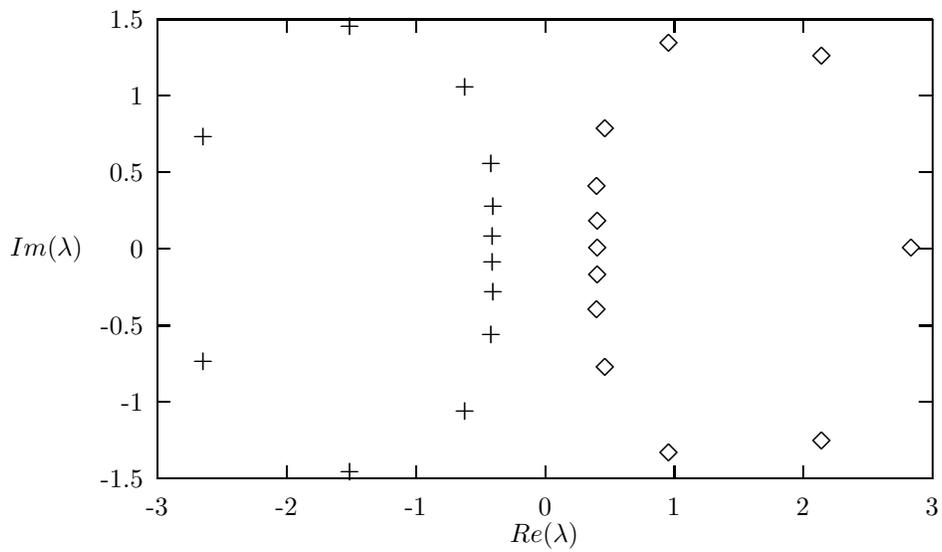
\begin{figure}
\input{fib-MZ.tex}
\caption{Eigenvalues of the Jacobi matrixes corresponding to
the RANLUX  generator with luxury level 0 (marked with pluses $+$).
The same for lagged Fibonacci generator with lags (24,10) (marked with
diamonds $\Diamond$).}
\label{E_L_0}
\end{figure}

The eigenvalues of RANLUX with luxury level $0$ are plotted on Fig~\ref{E_L_0}
together with the eigenvalues for corresponding lagged
Fibonacci generator with lags $(24,10)$. It is clear that eigenvalues
will collapse on a single curve changing the sign of $Re(\lambda)$.
These eigenvalues could be determined directly from the characteristic
polynomial for the productions rules, which reads as
\begin{equation}
\lambda_0^{24}-\lambda_0^{14}\pm 1=0
\label{mz-fib-pr}
\end{equation}
with plus sign for the Marsaglia-Zaman and minus sign for the
Fibonacci generator.
In Fig.~\ref{E_L_0} we have plotted $\lambda_0^{24}$
(we remind, that the dynamical system is obtained as the mapping of the
$r$-dimensional unit cube onto itself, after $p$ iterations of the
production rule. The luxury level 0 corresponds to $p=24$.)

\begin{figure}
\input{MZ-lux01.tex}
\caption{Eigenvalues of the Jacobi matrixes corresponding to
the RANLUX generator with luxury level 0 (pluses $+$) and to the RANLUX
generator with luxury  level 1 (diamonds $\Diamond$).}
\label{lux01}
\end{figure}
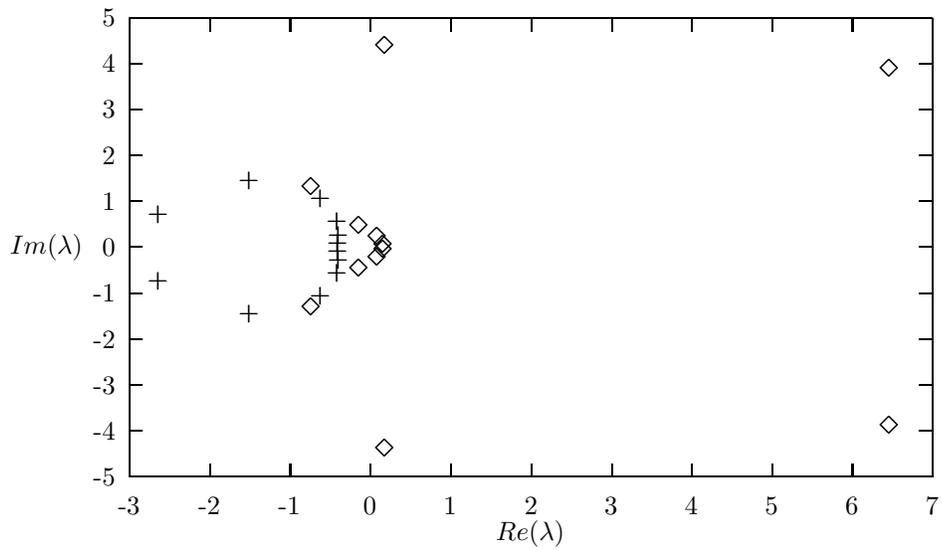

Fig.~\ref{lux01} demonstrates how the eigenvalues change going from the
luxury level 0 to the luxury level 1 in the case of the RANLUX generator.

Below we will produce data for the lagged Fibonacci generator with $p>r$
(we generate $p$ random numbers and use only the last $r$ of them)
taking into account the similarity in many aspects of the lagged Fibonacci
and the RANLUX generators described in this paper.

\begin{figure}
\epsfxsize=350pt
\epsffile{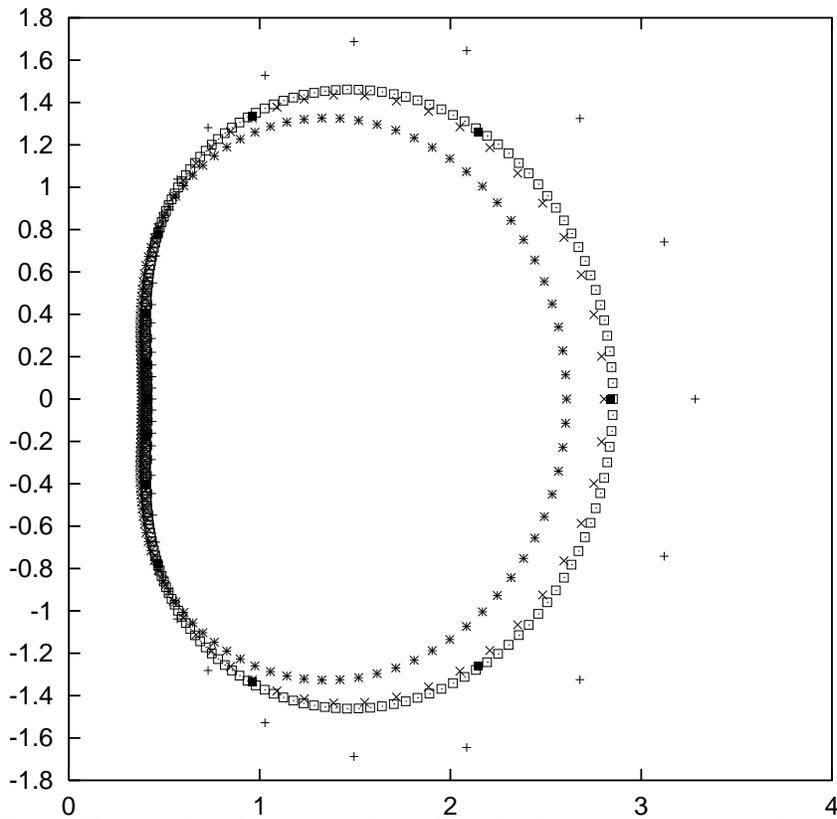}
\caption{The complex plane eigenvalues of the Jacobi matrices
corresponding to the lagged Fibonacci generator with lags (24,10),
(36,11),(89,38),(127,64) and (250,103)
indicated by $(\Box,+,\times,*,\Box)$ respectively.}
\label{fib-all}
\end{figure}

 First, we show how the eigenvalues depend on the value of the lag $r$.
Fig.~\ref{fib-all} shows, in particular, that the eigenvalues for (24,10)
and (250,103) lie almost on the same curve. We will plot below
the eigenvalues for the last case in order to have more detailed pictures.

\begin{figure}
\centering
\epsfxsize=120pt
\epsffile{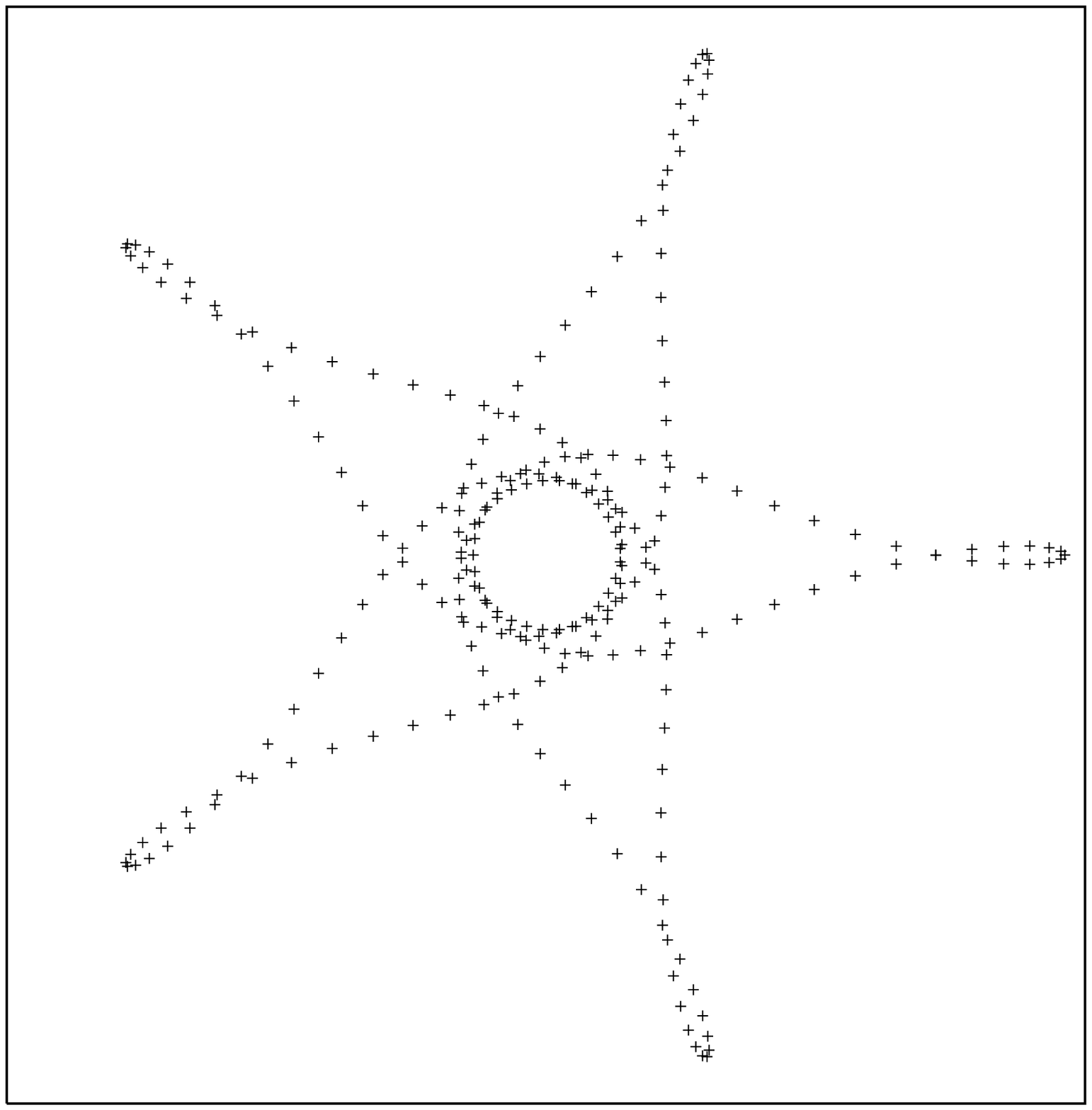}
\epsfxsize=120pt
\epsffile{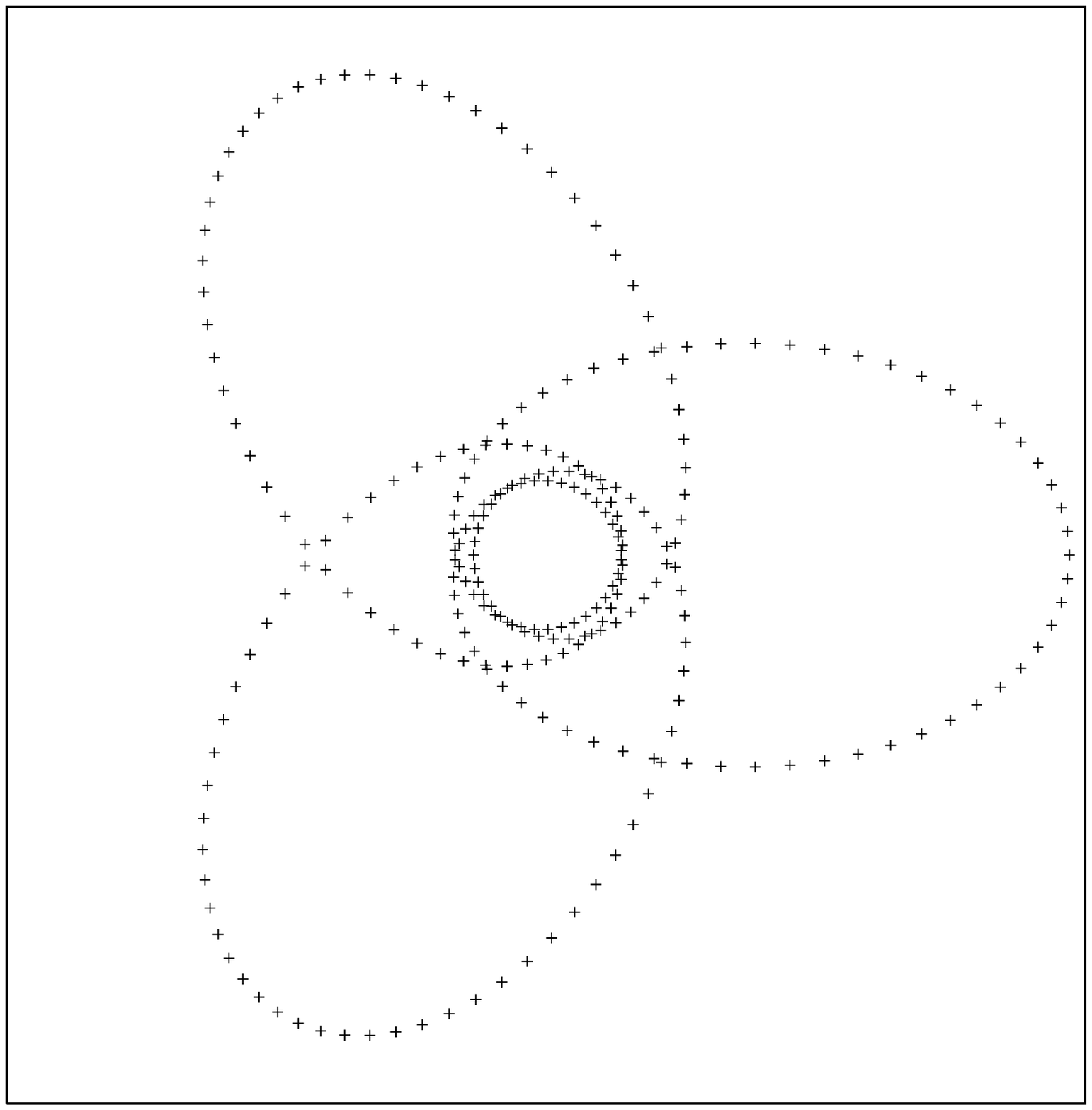}
\epsfxsize=120pt
\epsffile{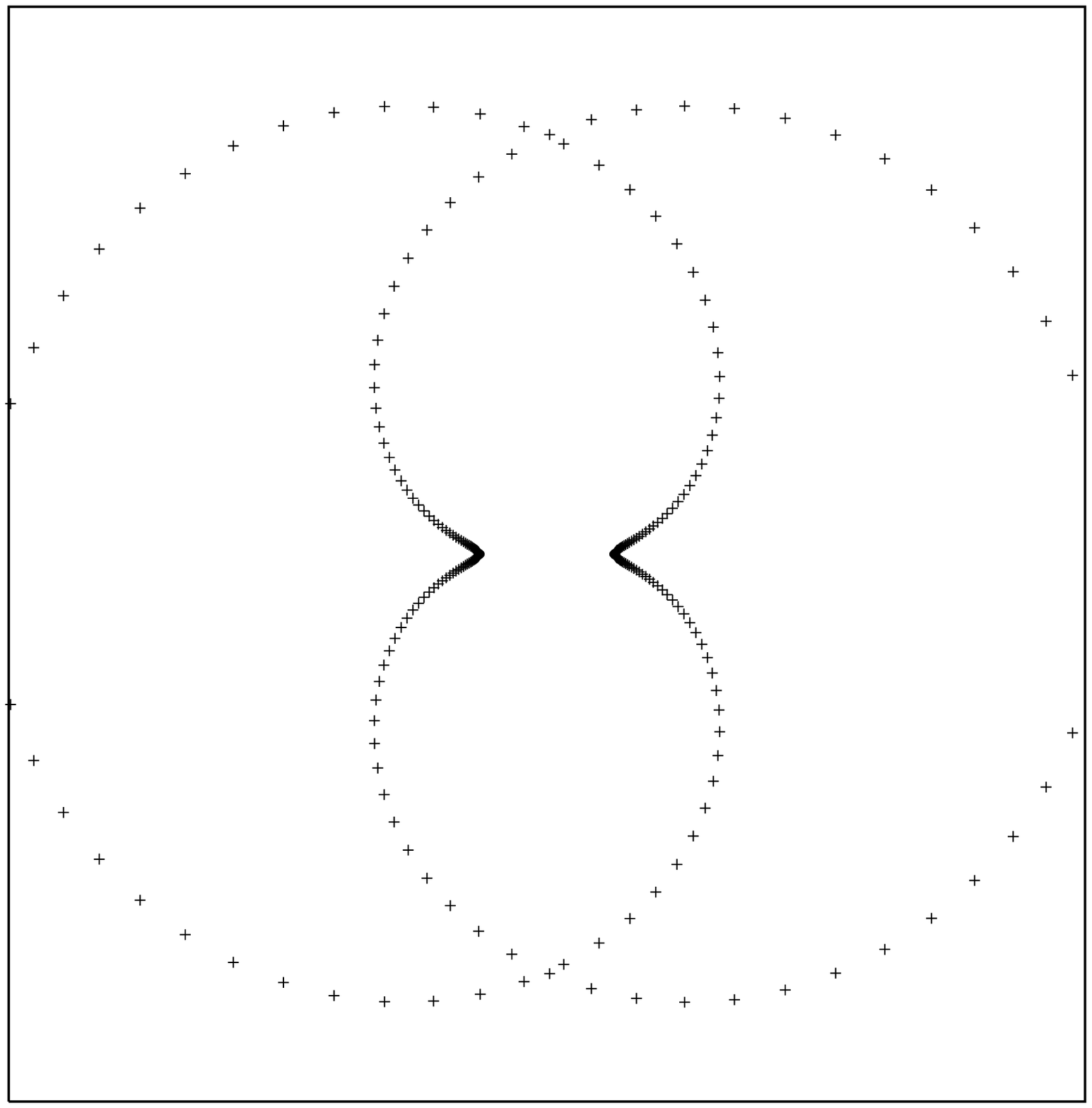}
\caption{Complex plane eigenvalues
for the lagged Fibonacci generator with lags (250,103).
The number of discarded PRN's is 3, 5 and 22 going from top to bottom.}
\label{zoo1}
\end{figure}

\begin{figure}
\epsfxsize=120pt
\epsffile{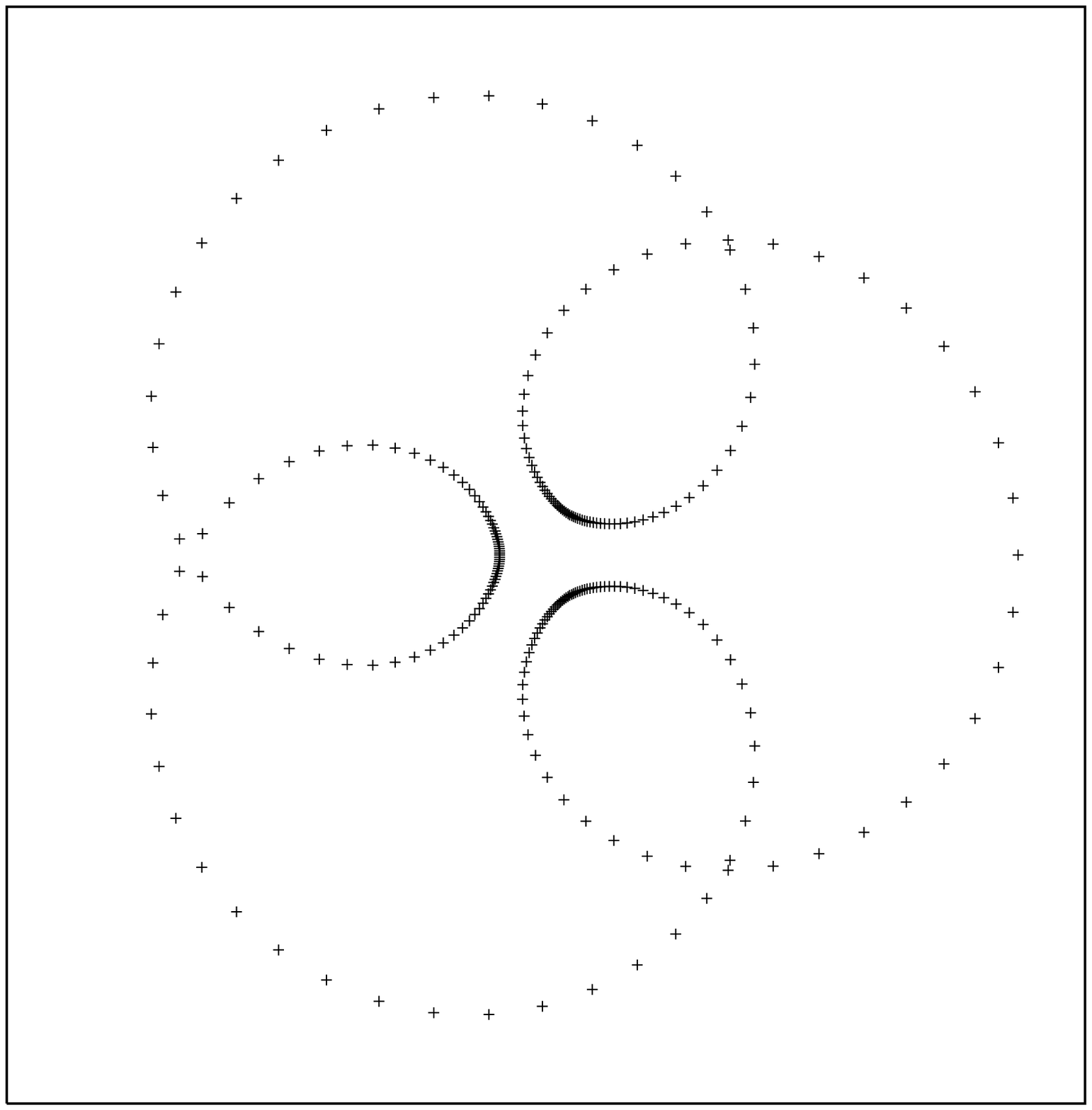}
\epsfxsize=120pt
\epsffile{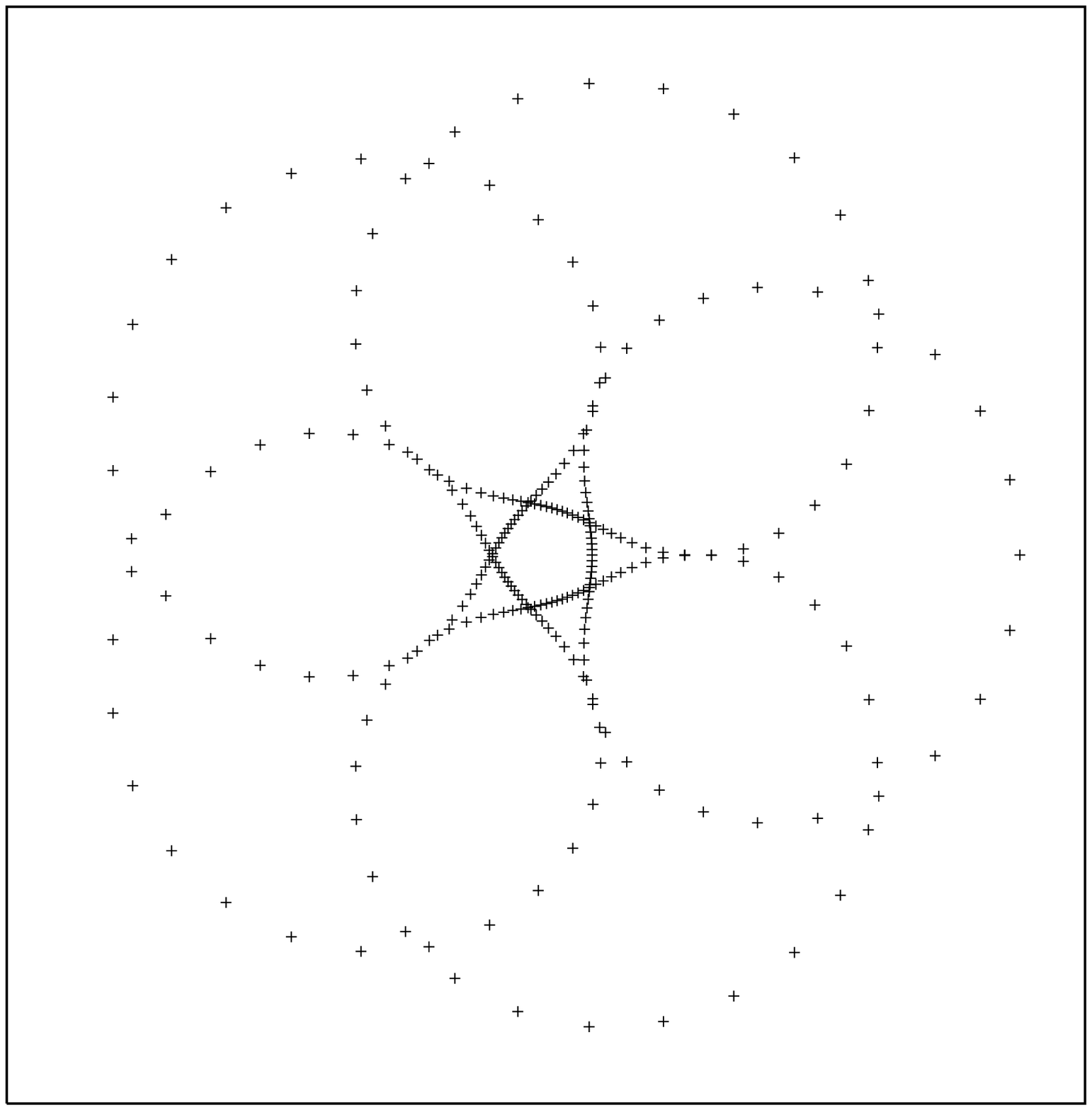}
\epsfxsize=120pt
\epsffile{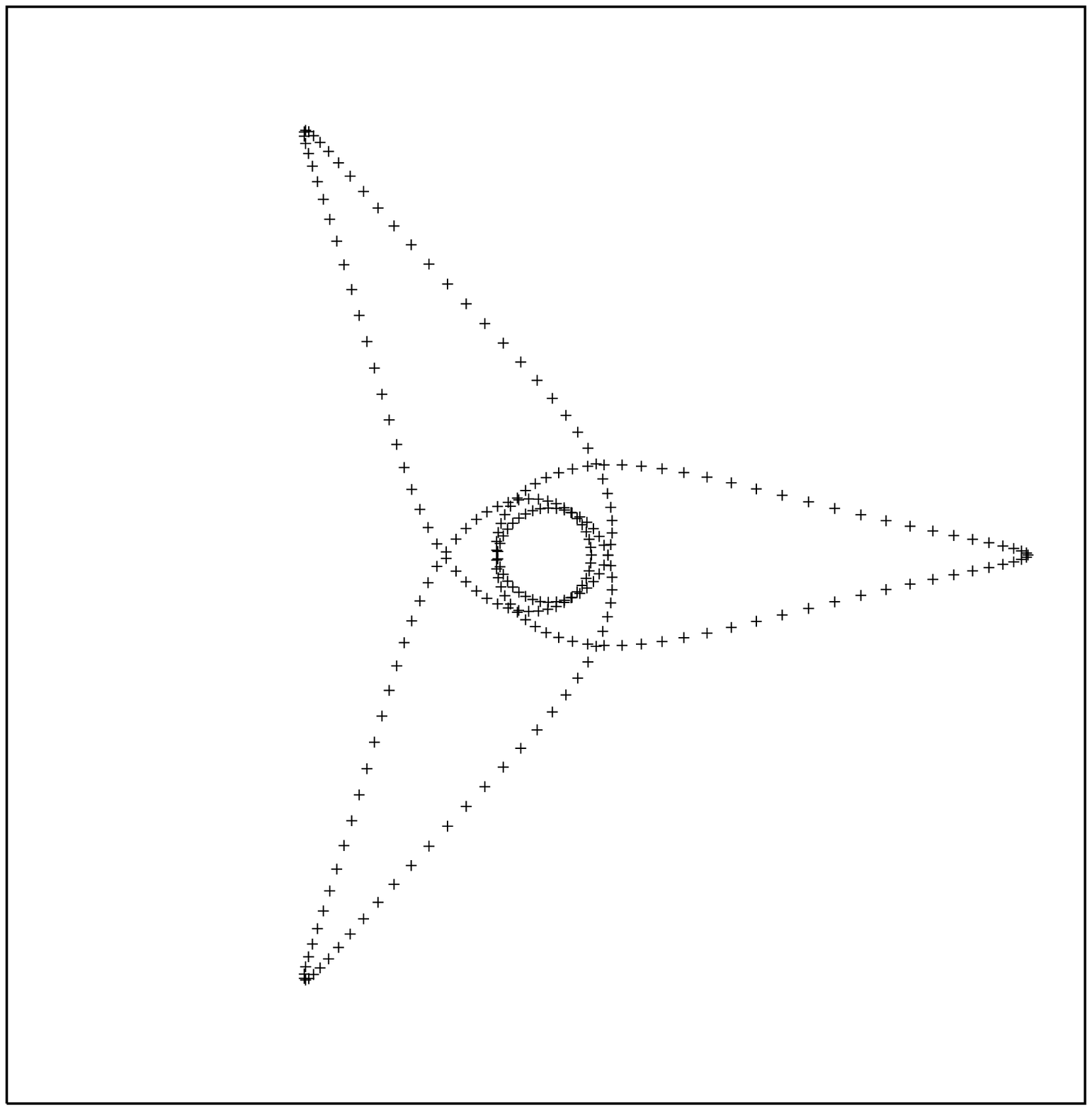}
\caption{Complex plane of eigenvalues
for the lagged Fibonacci generator with lags (250,103).
The number of discarded PRN's is 49, 50 and 54 going from top to bottom.}
\label{zoo2}
\end{figure}

\begin{figure}
\centering
\epsfxsize=120pt
\epsffile{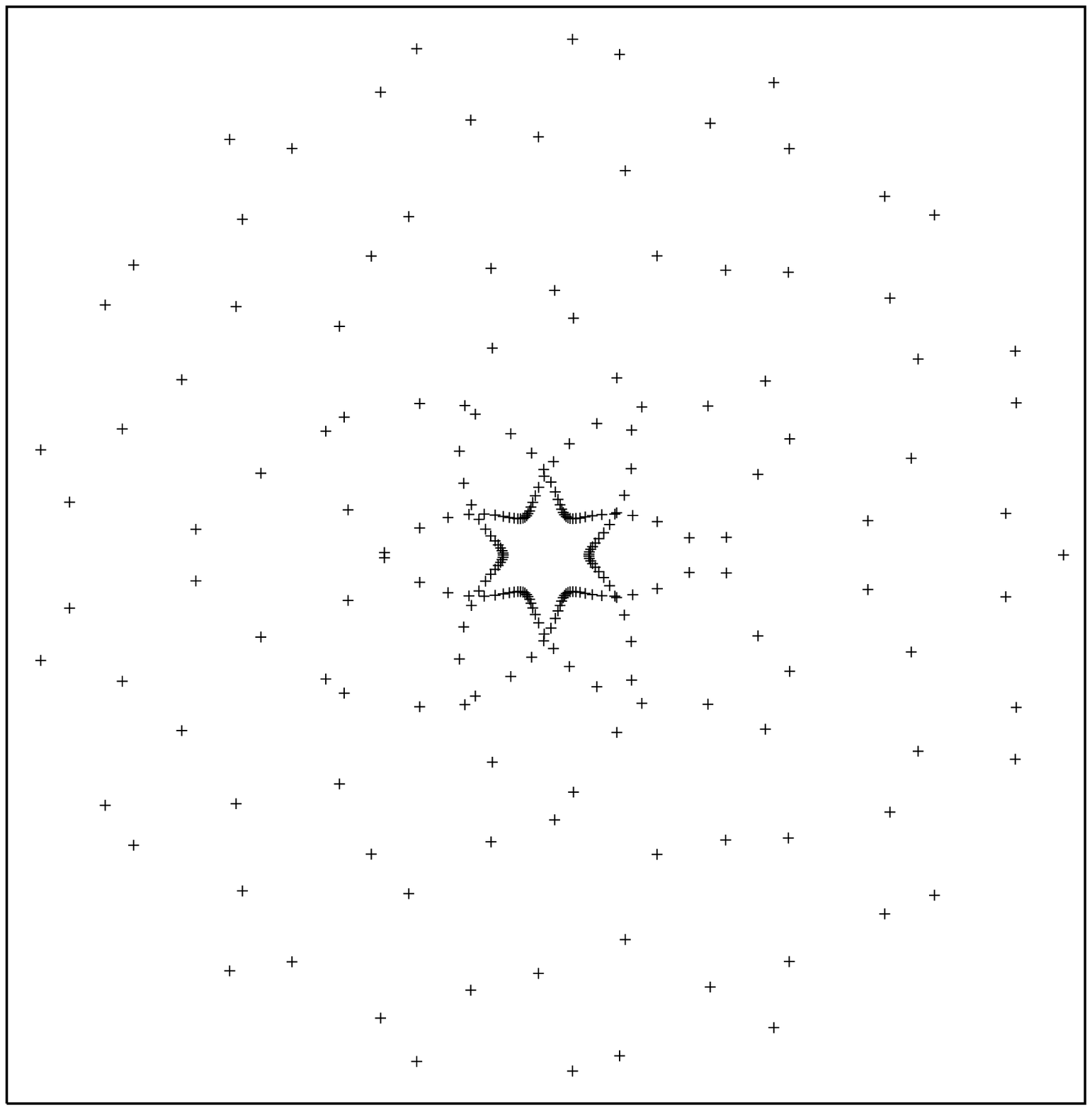}
\epsfxsize=120pt
\epsffile{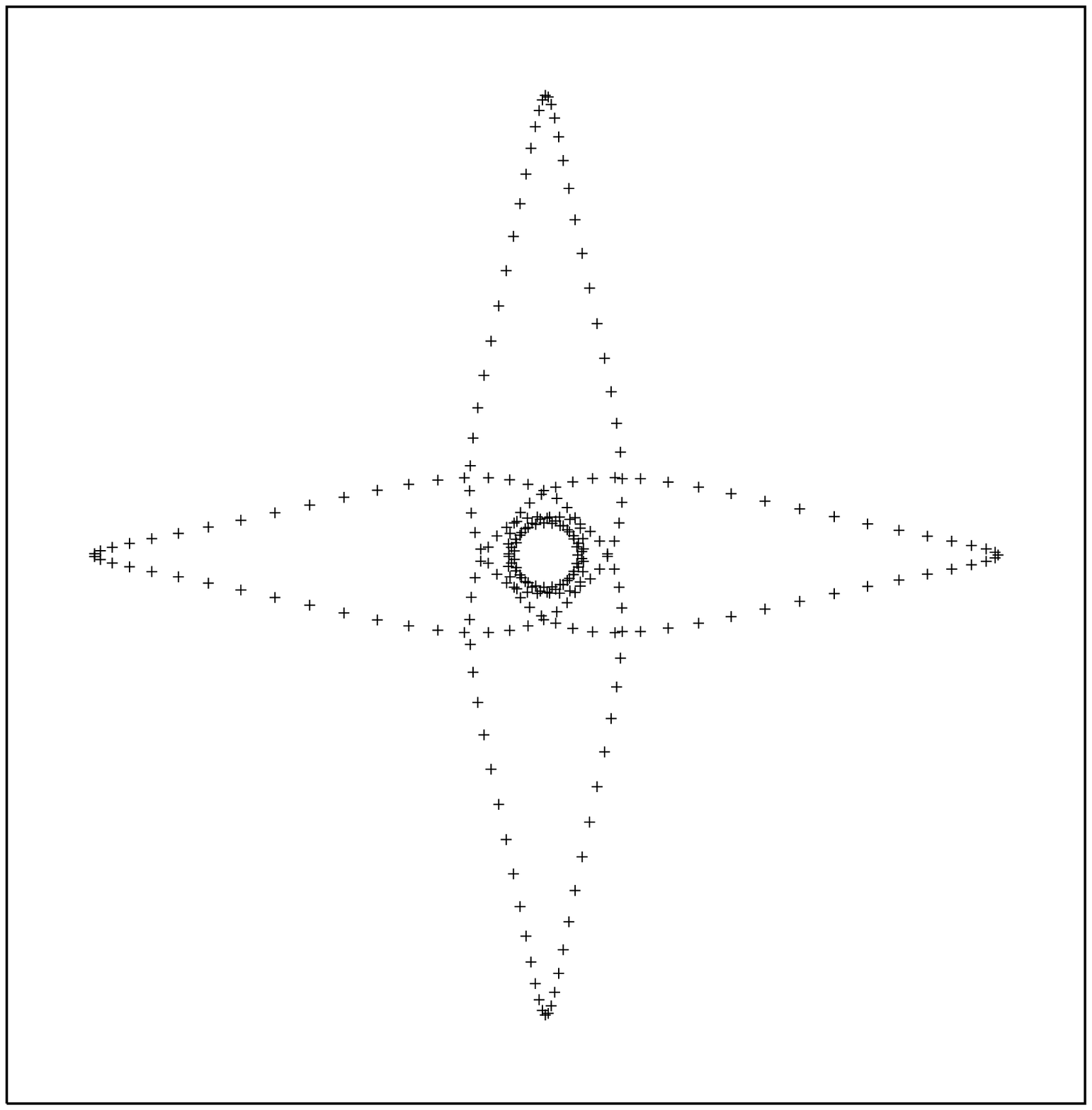}
\epsfxsize=120pt
\epsffile{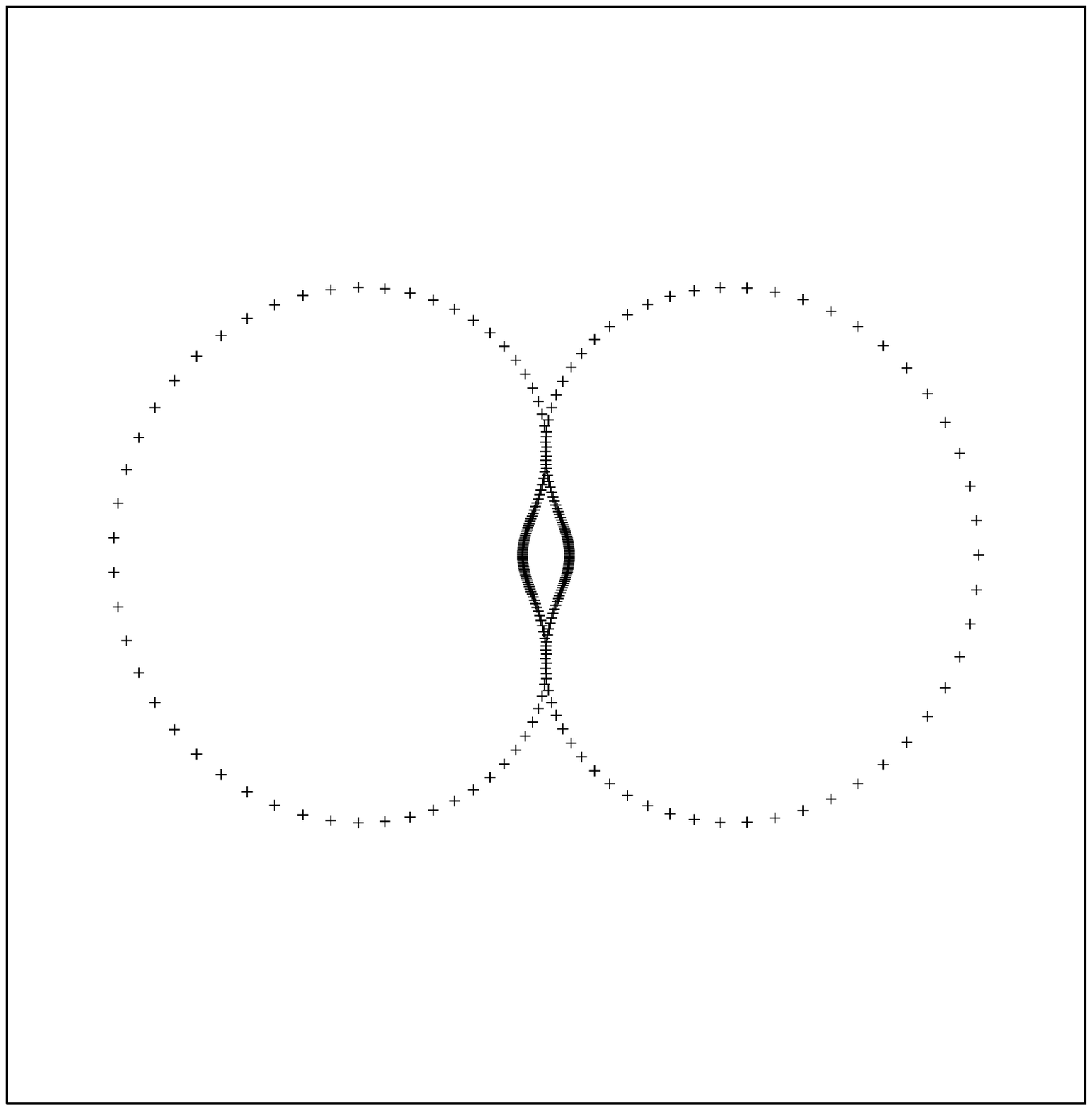}
\caption{Complex plane of eigenvalues
for the lagged Fibonacci generator with lags (250,103).
Number of discarded PRN's is 71, 92 and 125 going from top to bottom.}
\label{zoo3}
\end{figure}

\begin{figure}
\centering
\epsfxsize=120pt
\epsffile{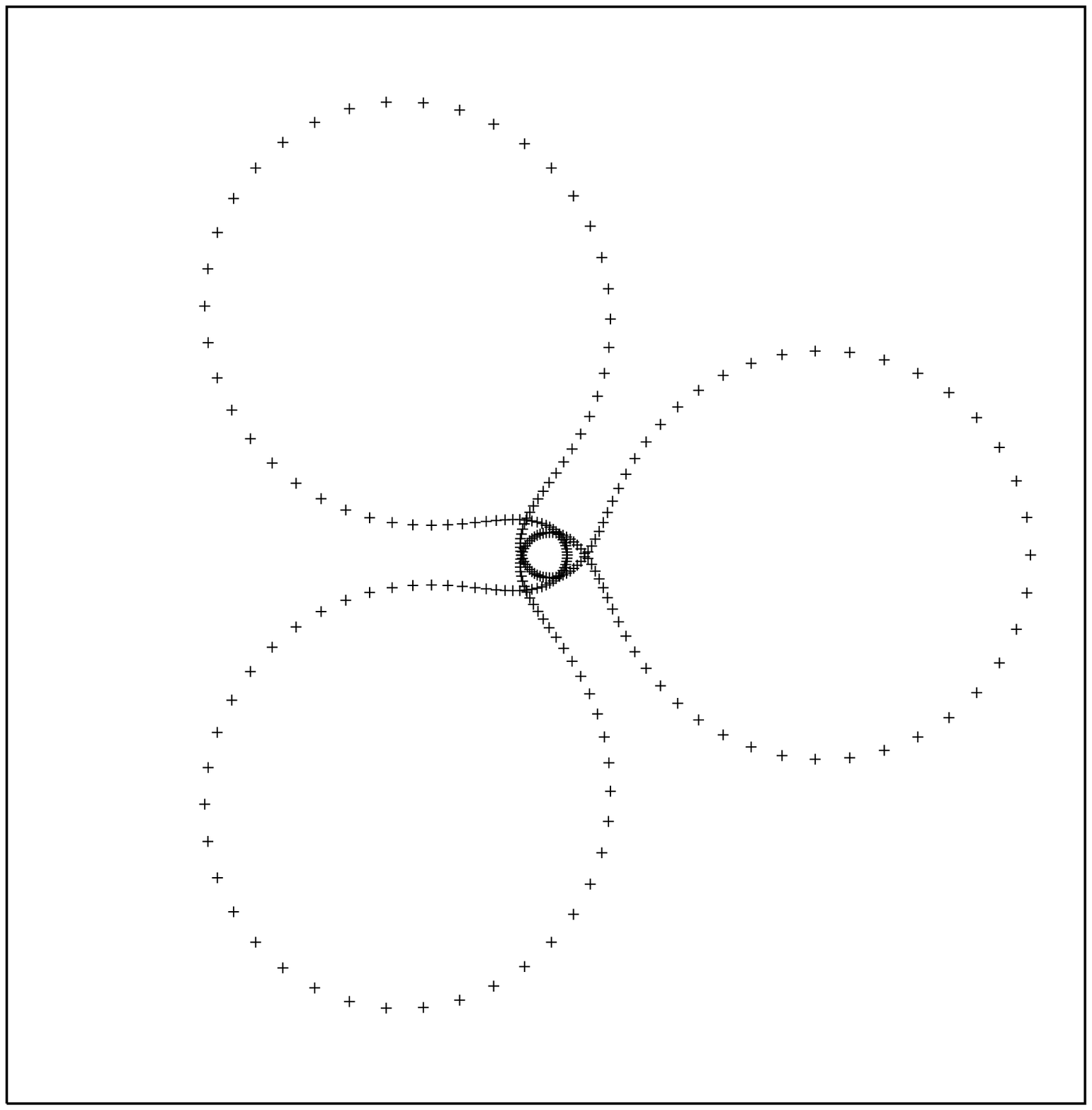}
\epsfxsize=120pt
\epsffile{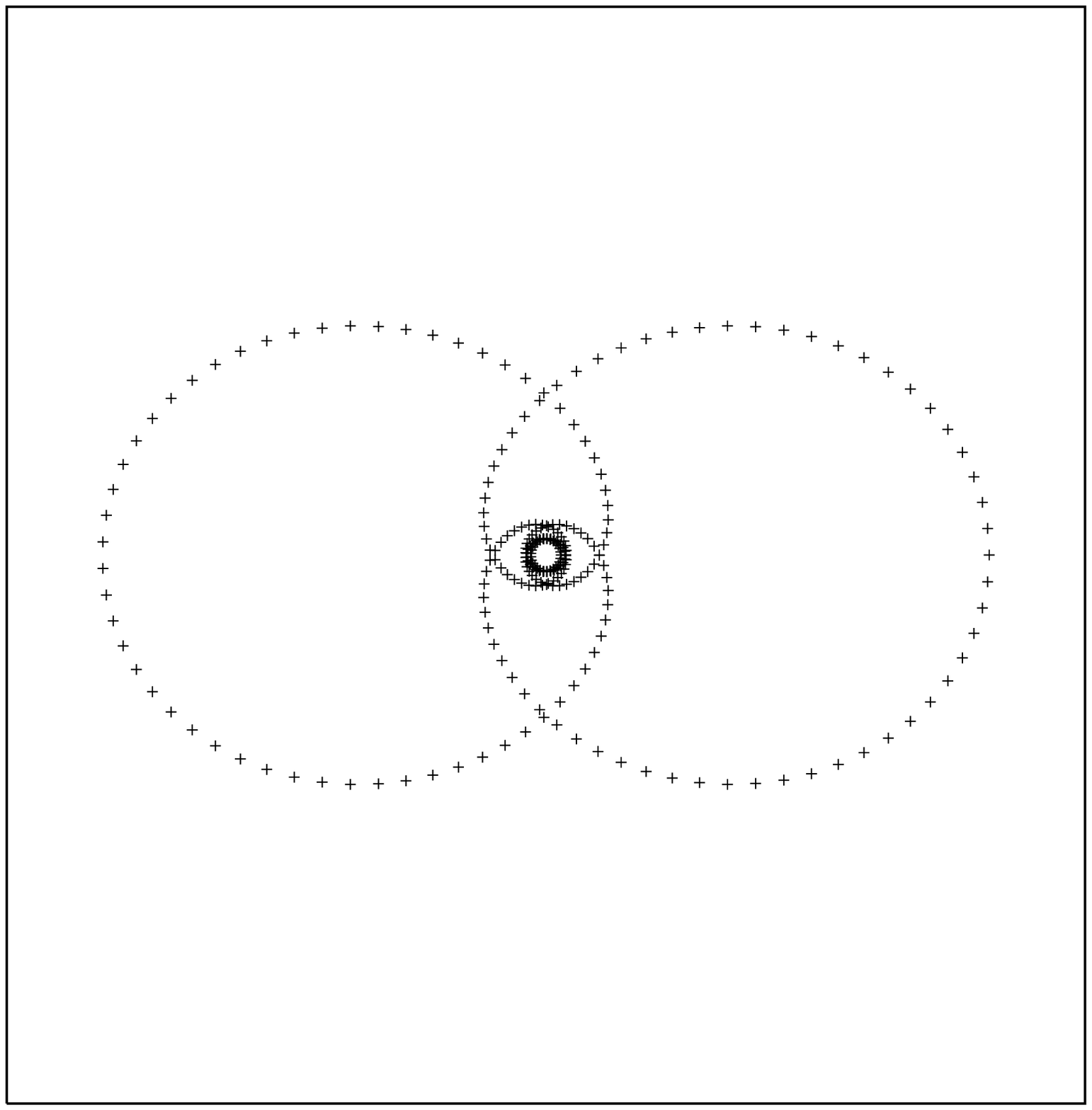}
\epsfxsize=120pt
\epsffile{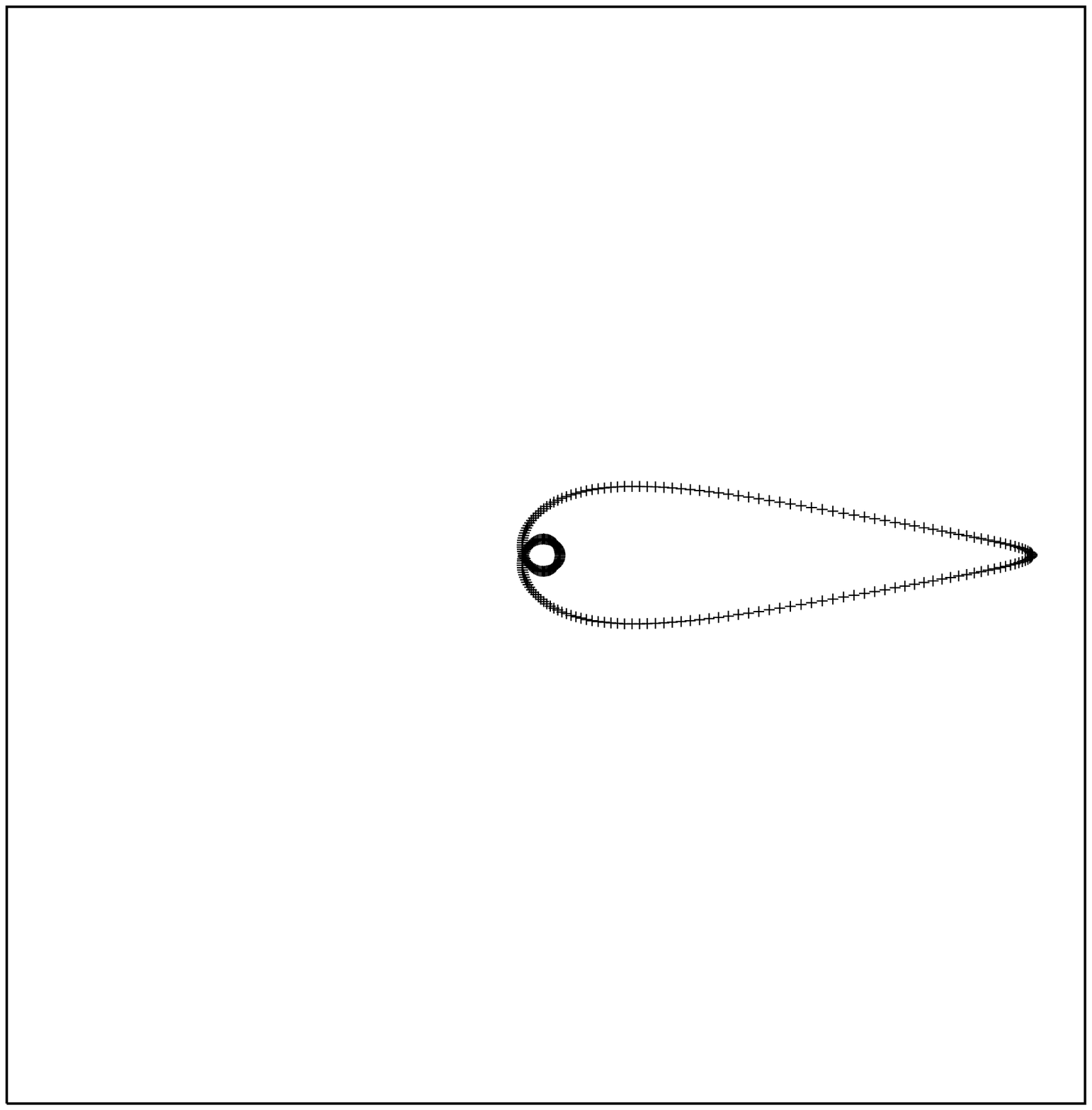}
\caption{Complex plane of eigenvalues
for the lagged Fibonacci generator with lags (250,103).
The number of discarded PRN's is 152, 184 and 206 going from top to
bottom.}
\label{zoo4}
\end{figure}

 In Figs.\ref{zoo1}-\ref{zoo4} we show how the eigenvalues change
when the parameter $p$ is varied. One clearly sees that the
resonances in eigenvalues reflect the interplay of the lags $(r,s)$
and the parameter $p$.
These pictures reflect in some sense the complexity of the phase space
of our dynamical systems.

\section{Conclusions}
\label{sec-Con}

 In this paper we have shown that the sequences of random numbers
generated using both the lagged Fibonacci and the subtract-with-carry
generators have equivalent correlation properties, at
least as far as the random walk and the cluster algorithms are concerned.

 We have argued that correlations in random numbers on the lag distances
decrease with increase of the luxury levels, although not monotonically. 
There are also resonances when the number $p-r$ of discarded random
numbers coincides with linear combinations of lags. We also have
argued that correlations on these distances are more important
than correlations between "trajectories" of random numbers in phase space.

 Comparisons of lagged Fibonacci and shift register generators
have been presented in \cite{SHB} and our analysis demonstrates
that they behave in a qualitatively similar way.

 One can conclude, that all recipes using lags will produce sequences
of random numbers with the same qualitative features. It is important
that knowing these bad properties we can avoid possible deviations
either using the idea of luxury levels \cite{lush,james}, or decimated
sequences \cite{At,lh} or combinations of two or more random
number generators \cite{At,lh,spp3} or generators with a larger
number of lags \cite{At,Ziff}.

Therefore rephrasing Orwell: "All algorithms for generating random
numbers are equally bad, but some of them are more equal than others"
\cite{orwell}).

\section{Acknowledgments}

The authors are thankful to G. Marchesini whose reasonable question on
RANLUX initiated this work. Discussions and help from M. Comi, M.
Enriotti, S.A. Krashakov and G. Salam were very important.
LNS thanks to Theoretical Group of Milan University for the kind
hospitality and to Italian Ministery for Foreign Affairs for a Fellowship.
This work has also been partially supported
by grants from RFBR, INTAS and NWO.

\begin{table}
\caption{Comparison of probability deviations, computed using directed
random walk test with lagged Fibonacci (LF) and subtract-with-carry (SWC)
random number generators, and predicted by formulas \ref{dPr}
and \ref{dPr1}.}

\begin{tabular}{|l|ll|ll|}
\label{Table1}
 & \multicolumn{2}{c|}{$\mu=31/32$} &
\multicolumn{2}{c|}{$\mu=15/16$} \\ \hline
\multicolumn{1}{|c}{} & \multicolumn{1}{c}{$\delta P(r)$} &
\multicolumn{1}{c|}{$\delta P(r+1)$} & \multicolumn{1}{c}{$\delta P(r)$}
& \multicolumn{1}{c|}{$\delta P(r+1)$} \\ \hline\hline
Expected & -0.48387...   & 0.03146...      & -0.4666...   & 0.06319... \\
LF   & -0.48304(55)  & 0.03221(81)     & -0.46636(47) & 0.06290(85) \\
SWC   & -0.48299(64)  & 0.03122(81)     & -0.46696(63) & 0.06243(113) \\
\end{tabular}
\end{table}

\newpage

\newpage

\end{document}

%% file: fib-MZ.tex
\setlength{\unitlength}{0.240900pt}
\ifx\plotpoint\undefined\newsavebox{\plotpoint}\fi
\sbox{\plotpoint}{\rule[-0.200pt]{0.400pt}{0.400pt}}%
\begin{picture}(1500,900)(0,0)
\font\gnuplot=cmr10 at 10pt
\gnuplot
\sbox{\plotpoint}{\rule[-0.200pt]{0.400pt}{0.400pt}}%
\put(219.0,134.0){\rule[-0.200pt]{4.818pt}{0.400pt}}
\put(197,134){\makebox(0,0)[r]{-1.5}}
\put(1416.0,134.0){\rule[-0.200pt]{4.818pt}{0.400pt}}
\put(219.0,254.0){\rule[-0.200pt]{4.818pt}{0.400pt}}
\put(197,254){\makebox(0,0)[r]{-1}}
\put(1416.0,254.0){\rule[-0.200pt]{4.818pt}{0.400pt}}
\put(219.0,374.0){\rule[-0.200pt]{4.818pt}{0.400pt}}
\put(197,374){\makebox(0,0)[r]{-0.5}}
\put(1416.0,374.0){\rule[-0.200pt]{4.818pt}{0.400pt}}
\put(219.0,495.0){\rule[-0.200pt]{4.818pt}{0.400pt}}
\put(197,495){\makebox(0,0)[r]{0}}
\put(1416.0,495.0){\rule[-0.200pt]{4.818pt}{0.400pt}}
\put(219.0,615.0){\rule[-0.200pt]{4.818pt}{0.400pt}}
\put(197,615){\makebox(0,0)[r]{0.5}}
\put(1416.0,615.0){\rule[-0.200pt]{4.818pt}{0.400pt}}
\put(219.0,735.0){\rule[-0.200pt]{4.818pt}{0.400pt}}
\put(197,735){\makebox(0,0)[r]{1}}
\put(1416.0,735.0){\rule[-0.200pt]{4.818pt}{0.400pt}}
\put(219.0,855.0){\rule[-0.200pt]{4.818pt}{0.400pt}}
\put(197,855){\makebox(0,0)[r]{1.5}}
\put(1416.0,855.0){\rule[-0.200pt]{4.818pt}{0.400pt}}
\put(219.0,134.0){\rule[-0.200pt]{0.400pt}{4.818pt}}
\put(219,89){\makebox(0,0){-3}}
\put(219.0,835.0){\rule[-0.200pt]{0.400pt}{4.818pt}}
\put(422.0,134.0){\rule[-0.200pt]{0.400pt}{4.818pt}}
\put(422,89){\makebox(0,0){-2}}
\put(422.0,835.0){\rule[-0.200pt]{0.400pt}{4.818pt}}
\put(625.0,134.0){\rule[-0.200pt]{0.400pt}{4.818pt}}
\put(625,89){\makebox(0,0){-1}}
\put(625.0,835.0){\rule[-0.200pt]{0.400pt}{4.818pt}}
\put(828.0,134.0){\rule[-0.200pt]{0.400pt}{4.818pt}}
\put(828,89){\makebox(0,0){0}}
\put(828.0,835.0){\rule[-0.200pt]{0.400pt}{4.818pt}}
\put(1030.0,134.0){\rule[-0.200pt]{0.400pt}{4.818pt}}
\put(1030,89){\makebox(0,0){1}}
\put(1030.0,835.0){\rule[-0.200pt]{0.400pt}{4.818pt}}
\put(1233.0,134.0){\rule[-0.200pt]{0.400pt}{4.818pt}}
\put(1233,89){\makebox(0,0){2}}
\put(1233.0,835.0){\rule[-0.200pt]{0.400pt}{4.818pt}}
\put(1436.0,134.0){\rule[-0.200pt]{0.400pt}{4.818pt}}
\put(1436,89){\makebox(0,0){3}}
\put(1436.0,835.0){\rule[-0.200pt]{0.400pt}{4.818pt}}
\put(219.0,134.0){\rule[-0.200pt]{293.175pt}{0.400pt}}
\put(1436.0,134.0){\rule[-0.200pt]{0.400pt}{173.689pt}}
\put(219.0,855.0){\rule[-0.200pt]{293.175pt}{0.400pt}}
\put(45,494){\makebox(0,0){$Im(\lambda)$}}
\put(827,44){\makebox(0,0){$Re(\lambda)$}}
\put(219.0,134.0){\rule[-0.200pt]{0.400pt}{173.689pt}}
\put(1403,495){\raisebox{-.8pt}{\makebox(0,0){$\Diamond$}}}
\put(1262,797){\raisebox{-.8pt}{\makebox(0,0){$\Diamond$}}}
\put(1262,192){\raisebox{-.8pt}{\makebox(0,0){$\Diamond$}}}
\put(1022,816){\raisebox{-.8pt}{\makebox(0,0){$\Diamond$}}}
\put(1022,173){\raisebox{-.8pt}{\makebox(0,0){$\Diamond$}}}
\put(922,682){\raisebox{-.8pt}{\makebox(0,0){$\Diamond$}}}
\put(922,307){\raisebox{-.8pt}{\makebox(0,0){$\Diamond$}}}
\put(909,591){\raisebox{-.8pt}{\makebox(0,0){$\Diamond$}}}
\put(909,398){\raisebox{-.8pt}{\makebox(0,0){$\Diamond$}}}
\put(910,495){\raisebox{-.8pt}{\makebox(0,0){$\Diamond$}}}
\put(910,537){\raisebox{-.8pt}{\makebox(0,0){$\Diamond$}}}
\put(910,452){\raisebox{-.8pt}{\makebox(0,0){$\Diamond$}}}
\put(1403,495){\raisebox{-.8pt}{\makebox(0,0){$\Diamond$}}}
\put(1262,797){\raisebox{-.8pt}{\makebox(0,0){$\Diamond$}}}
\put(1262,192){\raisebox{-.8pt}{\makebox(0,0){$\Diamond$}}}
\put(1022,816){\raisebox{-.8pt}{\makebox(0,0){$\Diamond$}}}
\put(1022,173){\raisebox{-.8pt}{\makebox(0,0){$\Diamond$}}}
\put(922,682){\raisebox{-.8pt}{\makebox(0,0){$\Diamond$}}}
\put(922,307){\raisebox{-.8pt}{\makebox(0,0){$\Diamond$}}}
\put(909,591){\raisebox{-.8pt}{\makebox(0,0){$\Diamond$}}}
\put(909,398){\raisebox{-.8pt}{\makebox(0,0){$\Diamond$}}}
\put(910,495){\raisebox{-.8pt}{\makebox(0,0){$\Diamond$}}}
\put(910,537){\raisebox{-.8pt}{\makebox(0,0){$\Diamond$}}}
\put(910,452){\raisebox{-.8pt}{\makebox(0,0){$\Diamond$}}}
\put(702,749){\makebox(0,0){$+$}}
\put(746,428){\makebox(0,0){$+$}}
\put(291,671){\makebox(0,0){$+$}}
\put(745,515){\makebox(0,0){$+$}}
\put(521,145){\makebox(0,0){$+$}}
\put(743,629){\makebox(0,0){$+$}}
\put(702,749){\makebox(0,0){$+$}}
\put(746,428){\makebox(0,0){$+$}}
\put(291,671){\makebox(0,0){$+$}}
\put(745,515){\makebox(0,0){$+$}}
\put(521,145){\makebox(0,0){$+$}}
\put(743,629){\makebox(0,0){$+$}}
\put(702,240){\makebox(0,0){$+$}}
\put(746,561){\makebox(0,0){$+$}}
\put(291,318){\makebox(0,0){$+$}}
\put(745,474){\makebox(0,0){$+$}}
\put(521,844){\makebox(0,0){$+$}}
\put(743,360){\makebox(0,0){$+$}}
\put(702,240){\makebox(0,0){$+$}}
\put(746,561){\makebox(0,0){$+$}}
\put(291,318){\makebox(0,0){$+$}}
\put(745,474){\makebox(0,0){$+$}}
\put(521,844){\makebox(0,0){$+$}}
\put(743,360){\makebox(0,0){$+$}}
\end{picture}

%% file: MZ-lux01.tex
\setlength{\unitlength}{0.240900pt}
\ifx\plotpoint\undefined\newsavebox{\plotpoint}\fi
\sbox{\plotpoint}{\rule[-0.200pt]{0.400pt}{0.400pt}}%
\begin{picture}(1500,900)(0,0)
\font\gnuplot=cmr10 at 10pt
\gnuplot
\sbox{\plotpoint}{\rule[-0.200pt]{0.400pt}{0.400pt}}%
\put(175.0,134.0){\rule[-0.200pt]{4.818pt}{0.400pt}}
\put(153,134){\makebox(0,0)[r]{-5}}
\put(1416.0,134.0){\rule[-0.200pt]{4.818pt}{0.400pt}}
\put(175.0,206.0){\rule[-0.200pt]{4.818pt}{0.400pt}}
\put(153,206){\makebox(0,0)[r]{-4}}
\put(1416.0,206.0){\rule[-0.200pt]{4.818pt}{0.400pt}}
\put(175.0,278.0){\rule[-0.200pt]{4.818pt}{0.400pt}}
\put(153,278){\makebox(0,0)[r]{-3}}
\put(1416.0,278.0){\rule[-0.200pt]{4.818pt}{0.400pt}}
\put(175.0,350.0){\rule[-0.200pt]{4.818pt}{0.400pt}}
\put(153,350){\makebox(0,0)[r]{-2}}
\put(1416.0,350.0){\rule[-0.200pt]{4.818pt}{0.400pt}}
\put(175.0,422.0){\rule[-0.200pt]{4.818pt}{0.400pt}}
\put(153,422){\makebox(0,0)[r]{-1}}
\put(1416.0,422.0){\rule[-0.200pt]{4.818pt}{0.400pt}}
\put(175.0,495.0){\rule[-0.200pt]{4.818pt}{0.400pt}}
\put(153,495){\makebox(0,0)[r]{0}}
\put(1416.0,495.0){\rule[-0.200pt]{4.818pt}{0.400pt}}
\put(175.0,567.0){\rule[-0.200pt]{4.818pt}{0.400pt}}
\put(153,567){\makebox(0,0)[r]{1}}
\put(1416.0,567.0){\rule[-0.200pt]{4.818pt}{0.400pt}}
\put(175.0,639.0){\rule[-0.200pt]{4.818pt}{0.400pt}}
\put(153,639){\makebox(0,0)[r]{2}}
\put(1416.0,639.0){\rule[-0.200pt]{4.818pt}{0.400pt}}
\put(175.0,711.0){\rule[-0.200pt]{4.818pt}{0.400pt}}
\put(153,711){\makebox(0,0)[r]{3}}
\put(1416.0,711.0){\rule[-0.200pt]{4.818pt}{0.400pt}}
\put(175.0,783.0){\rule[-0.200pt]{4.818pt}{0.400pt}}
\put(153,783){\makebox(0,0)[r]{4}}
\put(1416.0,783.0){\rule[-0.200pt]{4.818pt}{0.400pt}}
\put(175.0,855.0){\rule[-0.200pt]{4.818pt}{0.400pt}}
\put(153,855){\makebox(0,0)[r]{5}}
\put(1416.0,855.0){\rule[-0.200pt]{4.818pt}{0.400pt}}
\put(175.0,134.0){\rule[-0.200pt]{0.400pt}{4.818pt}}
\put(175,89){\makebox(0,0){-3}}
\put(175.0,835.0){\rule[-0.200pt]{0.400pt}{4.818pt}}
\put(301.0,134.0){\rule[-0.200pt]{0.400pt}{4.818pt}}
\put(301,89){\makebox(0,0){-2}}
\put(301.0,835.0){\rule[-0.200pt]{0.400pt}{4.818pt}}
\put(427.0,134.0){\rule[-0.200pt]{0.400pt}{4.818pt}}
\put(427,89){\makebox(0,0){-1}}
\put(427.0,835.0){\rule[-0.200pt]{0.400pt}{4.818pt}}
\put(553.0,134.0){\rule[-0.200pt]{0.400pt}{4.818pt}}
\put(553,89){\makebox(0,0){0}}
\put(553.0,835.0){\rule[-0.200pt]{0.400pt}{4.818pt}}
\put(679.0,134.0){\rule[-0.200pt]{0.400pt}{4.818pt}}
\put(679,89){\makebox(0,0){1}}
\put(679.0,835.0){\rule[-0.200pt]{0.400pt}{4.818pt}}
\put(806.0,134.0){\rule[-0.200pt]{0.400pt}{4.818pt}}
\put(806,89){\makebox(0,0){2}}
\put(806.0,835.0){\rule[-0.200pt]{0.400pt}{4.818pt}}
\put(932.0,134.0){\rule[-0.200pt]{0.400pt}{4.818pt}}
\put(932,89){\makebox(0,0){3}}
\put(932.0,835.0){\rule[-0.200pt]{0.400pt}{4.818pt}}
\put(1058.0,134.0){\rule[-0.200pt]{0.400pt}{4.818pt}}
\put(1058,89){\makebox(0,0){4}}
\put(1058.0,835.0){\rule[-0.200pt]{0.400pt}{4.818pt}}
\put(1184.0,134.0){\rule[-0.200pt]{0.400pt}{4.818pt}}
\put(1184,89){\makebox(0,0){5}}
\put(1184.0,835.0){\rule[-0.200pt]{0.400pt}{4.818pt}}
\put(1310.0,134.0){\rule[-0.200pt]{0.400pt}{4.818pt}}
\put(1310,89){\makebox(0,0){6}}
\put(1310.0,835.0){\rule[-0.200pt]{0.400pt}{4.818pt}}
\put(1436.0,134.0){\rule[-0.200pt]{0.400pt}{4.818pt}}
\put(1436,89){\makebox(0,0){7}}
\put(1436.0,835.0){\rule[-0.200pt]{0.400pt}{4.818pt}}
\put(175.0,134.0){\rule[-0.200pt]{303.775pt}{0.400pt}}
\put(1436.0,134.0){\rule[-0.200pt]{0.400pt}{173.689pt}}
\put(175.0,855.0){\rule[-0.200pt]{303.775pt}{0.400pt}}
\put(45,494){\makebox(0,0){$Im(\lambda)$}}
\put(805,44){\makebox(0,0){$Re(\lambda)$}}
\put(175.0,134.0){\rule[-0.200pt]{0.400pt}{173.689pt}}
\put(460,400){\raisebox{-.8pt}{\makebox(0,0){$\Diamond$}}}
\put(564,511){\raisebox{-.8pt}{\makebox(0,0){$\Diamond$}}}
\put(1368,214){\raisebox{-.8pt}{\makebox(0,0){$\Diamond$}}}
\put(573,490){\raisebox{-.8pt}{\makebox(0,0){$\Diamond$}}}
\put(576,811){\raisebox{-.8pt}{\makebox(0,0){$\Diamond$}}}
\put(535,461){\raisebox{-.8pt}{\makebox(0,0){$\Diamond$}}}
\put(460,400){\raisebox{-.8pt}{\makebox(0,0){$\Diamond$}}}
\put(564,511){\raisebox{-.8pt}{\makebox(0,0){$\Diamond$}}}
\put(1368,214){\raisebox{-.8pt}{\makebox(0,0){$\Diamond$}}}
\put(573,490){\raisebox{-.8pt}{\makebox(0,0){$\Diamond$}}}
\put(576,811){\raisebox{-.8pt}{\makebox(0,0){$\Diamond$}}}
\put(535,461){\raisebox{-.8pt}{\makebox(0,0){$\Diamond$}}}
\put(460,589){\raisebox{-.8pt}{\makebox(0,0){$\Diamond$}}}
\put(564,478){\raisebox{-.8pt}{\makebox(0,0){$\Diamond$}}}
\put(1368,775){\raisebox{-.8pt}{\makebox(0,0){$\Diamond$}}}
\put(573,499){\raisebox{-.8pt}{\makebox(0,0){$\Diamond$}}}
\put(576,178){\raisebox{-.8pt}{\makebox(0,0){$\Diamond$}}}
\put(535,528){\raisebox{-.8pt}{\makebox(0,0){$\Diamond$}}}
\put(460,589){\raisebox{-.8pt}{\makebox(0,0){$\Diamond$}}}
\put(564,478){\raisebox{-.8pt}{\makebox(0,0){$\Diamond$}}}
\put(1368,775){\raisebox{-.8pt}{\makebox(0,0){$\Diamond$}}}
\put(573,499){\raisebox{-.8pt}{\makebox(0,0){$\Diamond$}}}
\put(576,178){\raisebox{-.8pt}{\makebox(0,0){$\Diamond$}}}
\put(535,528){\raisebox{-.8pt}{\makebox(0,0){$\Diamond$}}}
\put(475,571){\makebox(0,0){$+$}}
\put(503,475){\makebox(0,0){$+$}}
\put(220,547){\makebox(0,0){$+$}}
\put(502,501){\makebox(0,0){$+$}}
\put(363,390){\makebox(0,0){$+$}}
\put(501,535){\makebox(0,0){$+$}}
\put(475,571){\makebox(0,0){$+$}}
\put(503,475){\makebox(0,0){$+$}}
\put(220,547){\makebox(0,0){$+$}}
\put(502,501){\makebox(0,0){$+$}}
\put(363,390){\makebox(0,0){$+$}}
\put(501,535){\makebox(0,0){$+$}}
\put(475,418){\makebox(0,0){$+$}}
\put(503,514){\makebox(0,0){$+$}}
\put(220,442){\makebox(0,0){$+$}}
\put(502,488){\makebox(0,0){$+$}}
\put(363,599){\makebox(0,0){$+$}}
\put(501,454){\makebox(0,0){$+$}}
\put(475,418){\makebox(0,0){$+$}}
\put(503,514){\makebox(0,0){$+$}}
\put(220,442){\makebox(0,0){$+$}}
\put(502,488){\makebox(0,0){$+$}}
\put(363,599){\makebox(0,0){$+$}}
\put(501,454){\makebox(0,0){$+$}}
\end{picture}